# The content and structure of dreams are coupled to affect

Luke Leckie[1,2*], Anya K. Bershad[3], Jes Heppler[4], Mason McClay[5], Sofiia Rappe[6], Jacob G. Foster[1,2,7*]

[1]Department of Informatics, Luddy School of Informatics, Computing, and Engineering, Indiana University; [2]Cognitive Science Program, Indiana University Bloomington; [3]Department of Psychiatry and Biobehavioral Sciences, Semel Institute for Neuroscience and Human Behavior, UCLA, Los Angeles, CA; [4]Department of Philosophy, University of California Berkeley, Berkeley, CA, [5]Department of Psychology, UCLA, Los Angeles, [6]Department of Philosophy II, Chair of Philosophy of Language and Cognition, Ruhr University Bochum, Germany; and [7]Santa Fe Institute.
*Corresponding authors: LL: lleckie@iu.edu; JGF: jacobf@iu.edu

## Abstract

Dreams offer a unique window into the cognitive and affective dynamics of the sleeping and the waking mind. Recent quantitative linguistic approaches have shown promise in obtaining corpus-level measures of dream sentiment and topic occurrence. However, it is currently unclear how the affective content of individual dreams relates to their semantic content and structure. Here, we combine word embedding, topic modeling, and network analysis to investigate this relationship. By applying Discourse Atom Topic Modeling (DATM) to the DreamBank corpus of >18K dream reports, we represent the latent themes arising within dream reports as a sparse dictionary of topics and identify the affective associations of those topics. We show that variation in dream report affect (valence and arousal) is associated with changes in topical content. By representing each dream report as a network of topics, we demonstrate that the affective content of dream narratives is also coupled to semantic structure. Positively valenced dream reports exhibit more coherent, structured, and linear narratives, whilst negatively valenced dreams have more narrative loops and dominant topics. Topic networks of high arousal dream reports are structurally dominated by few high arousal topics and incoherent topical connections, whereas low arousal dream reports contain more loops. These findings suggest that affective processes are associated with both the content and structure of dreams. Our approach showcases the potential of integrating natural language processing and network analysis with psychology to elucidate the interplay of affect, cognition and narrative in dreams. This methodology has broad applications for the study of narrated experience and psychiatric symptomatology.

## Significance statement

Dreams are windows into the mind with psychiatric importance. We integrate computational text and network analyses to study the relationship between emotional affect, content, and structure in >18K dream reports. We model the topics in dream reports to show that their thematic content changes with how positive and intense they are. By using networks to characterize the narrative progression of topics within dreams, we show that positive dreams follow linear, well-structured, narratives, whilst negative dreams contain many unrelated topics and narrative loops. Intense dreams also contain many unrelated topics, but have fewer loops than dull dreams. This research demonstrates the interrelation of affect, content, and structure in dreams and introduces tools to study this interplay in diverse report data.

## Introduction

Dreams are salient experiences that bear substantial cultural, personal, and clinical significance (1). The qualitative study of dream narratives as windows into the sleeping and waking mind has a rich history, spanning from ancient thinkers like Aristotle to the influential works of Freud and Jung in the 19th and 20th century (2–4). Some branches of contemporary psychotherapy still rely on dream analysis and many believe that dreams can serve as indicators of mental and physical health (1, 5). For example, nightmares, lucid dreaming, and a range of parasomnias have all been linked to particular physiological signals, such as distinct brain wave patterns (6–9). Moreover, disorders that can affect REM sleep, such as REM sleep behavior disorder, narcolepsy, and post-traumatic stress disorder, are often associated with more bizarre and negatively toned dream narratives (6, 10–12). In non-clinical populations, it is known that centers of emotion-processing within the brain experience heightened activation during REM sleep (13, 14) and that dream narratives are often dominated by stressors from one's waking life (15, 16). The nature of dreamt experience is, therefore, tightly linked to affect (the experience of emotion, mood, or feeling state).

Previous literature suggests that dream affect may be tied to the content and structure of dream narratives. For instance, Hartmann proposes that dreams are organised around a central image, or theme, that is a direct expression of a dreamer's dominant emotion (17, 18); by interconnecting this with other themes, individuals are able to recontextualise traumatic experiences within broader narrative structures, and so overcome them (18, 19). Similarly, Kramer proposes that dream narratives accomplish mood regulation by forming progressive sequences of new content into linear narratives, whilst emotional conflict and mood dysregulation manifest in dream narratives as repetitive patterns of traumatic content (20, 21). To understand the interrelation between affect, content, and structure within dream narratives quantitatively and at scale, it is essential to apply computational approaches to the study of dream reports.

In recent years, the scientific study of dream narratives has seen a resurgence, driven by improvements in computational tools such as natural language processing (NLP) (22). In particular, there has been increased interest in modeling the phenomenological content of dream reports through textual analysis and NLP (23–27). For example, analyses of dream word content have shown that the average affective rating of words in dream reports correlates with waking mood during the following morning (23). Additionally, by analysing the sequencing of words in dream reports, it has been shown that the dream reports of individuals with bipolar or obsessive-compulsive disorder exhibit more disconnected lexical structures (26, 27). However, most present NLP methods for characterising dream content —such as word frequency analysis, classical topic modeling, and lexicon-based sentiment scoring—are reliant upon computing the frequencies of words over entire collections of dream reports and are not tunable to represent the semantics or affect of words and topics as they occur within the specific context of dream reports (22, 25). Furthermore, current methods for characterising the linguistic structure of texts focus upon their low-level lexical composition, as derived from the sequencing of individual words, and do not encode the high-level topical flow of narratives (26, 27). Current computational approaches for analyzing dream report content and structure have, therefore, been unable to capture the discursive features that change within individual dream reports, such as the affective and semantic arcs that compose a narrative. Despite contemporary work making significant progress towards developing a quantitative understanding of dreamt experience, there is a present gap in our understanding of the dynamic interplay between the affective content, semantic content and semantic structure of dream reports.

In this study, we combine Discourse Atom Topic Modeling (DATM) (28) and network analysis to test the claim that the affective content of dream reports is associated with changes in topical content and semantic structure and, more specifically, that dream report valence (how positive or negative an experience is) is positively associated with increased narrative structure, coherence, and linearity, but is negatively associated with narrative fragmentation and disorganisation (17, 18, 21, 29, 30). We apply this approach to the DreamBank corpus (31) to show that the semantic content and semantic structure of dream reports are robustly associated with their affective content.

DreamBank is a publicly-available collection of over 18,000 dream reports sourced from a variety of individuals and research studies. This diverse corpus spans a wide age range and includes reports collected over a number of decades. By deploying a novel NLP toolkit to this corpus, we make several contributions towards advancing a quantitative understanding of dreams. First, by fitting word embedding and DATM to these dream reports, we are able to extract a rich representation of the thematic elements and affective meanings within dream reports. By using DATM, specifically, we are able to separate topics and whole narratives along two key affective dimensions–valence and arousal. Second, by comparing the prevalence of topics between dream reports scoring at the extremes of these dimensions, we show that the semantic content of dream reports changes along these affective dimensions. Third, we show that dream reports can be represented as networks of topics, thereby offering representations of semantic structure amenable to quantitative analysis. Finally, we use network analysis to reveal that structural properties of these topic networks are robustly associated with dream report valence and arousal. This study not only offers new insights into the relationships between dream report affect, content, and structure but also presents a quantitative approach that can be applied to the study of narrative in self-reports across various domains.

## Results

### Changes in dream report affective content are associated with changes in semantic content

To develop a quantitative understanding of dream reports, we deploy a natural language processing pipeline leveraging both word embedding and DATM to the DreamBank corpus. First, we train a word embedding on the corpus (32). Word embeddings provide a rich high-dimensional representation of corpora by mapping words to vectors within a continuous "semantic space". The distance between word vectors in the semantic space corresponds to their similarity in meaning: words with more similar contextual meanings are closer together and words with more dissimilar meanings are further apart from one another. Second, we assign topics to our embedding with DATM. DATM represents a semantic space as a sparse dictionary of "discourse atoms", or topics, which are mapped to new vectors within the same semantic space as words. As compared to other topic modeling approaches, such as LDA, DATM provides more interpretable topics, crisper topical boundaries, and topic-level embeddings (28).

By using DATM, we minimally represent dream report semantic space with only 130 topics (see SI Appendix for further details), which enables us to efficiently analyze the occurrence and structure of common thematic elements within dream reports.

We first examine how the affective meanings of dream report topics vary across two key dimensions: valence, which refers to how positive or negative an experience is, and arousal, which corresponds to the intensity of an experience. To do this, we use our embedding to create separate valence and arousal dimensions from standard seed words (33–36) representing opposing poles of each of these affective axes (positive valence seeds: good, happy, safe, confident, relaxed, love; negative valence seeds: bad, sad, scared, anxious, angry, disgust; high arousal seeds: energetic, delighted, excited, tense, angry, frustrated; low arousal seeds: lazy, depressed, tired, secure, relaxed, calm). We then compute topic-level scores of valence and arousal by loading each of our topics onto these dimensions. Topic affective loadings are measured by cosine similarity, which ranges between +1 and -1. Here, positive values of cosine similarity, closer to +1, indicate that topic has stronger contextual associations with positively valenced or higher arousal affective states, whilst more negative values, closer to -1, indicate a stronger contextual association with more negatively valenced or lower arousal affective states.

In our topic model, the most positively valenced topics relate to "dream characters and social acts" (topic 50) and "creative expression" (topic 54), such as playing and dancing. In contrast, the most negatively valenced topics relate to "negative emotions" (topic 121), such as fear and frustration, and "abrupt actions", such as waking-up from sleep or running (topic 120). Some of the highest arousal topics in our model relate to "conflict" (topic 75) and work (topic 74), whilst the lowest arousal topics relate to material descriptions of objects (topics 10 and 72) and "grammatical structure" (topic 64). Interestingly, we find several topics that relate directly to named dream characters (topics 20, 46, 50, 86 and 127), all with different affective associations, which highlights the ubiquitous and affective nature of dream characters in dream reports. We find no significant association between topic valence and arousal (Fig. 1A, Pearson correlation[95% CI low, 95% CI high]=-0.11[-0.28, 0.06], p=0.19).

Next, we investigate the relationship between the overall affective content of dream reports and the semantic content of dream reports. It should be noted that although affect is a component of semantics, it is only a fraction of it and the two are known to be analytically separable (37, 38). To do this, we first represent each narrative as a sequence of topics by sliding a context window over each narrative and inferring the corresponding topic for each window (See SI Appendix and (28) for further details). We then obtain dream-level scores of valence and arousal by computing median valence and arousal over the set of topics present within each dream narrative. We find that although most dream reports rate relatively neutrally for both valence and arousal, there is a negative correlation between dream-level valence and arousal (Fig. 1A, Pearson correlation=-0.33[-0.34, -0.32], p<0.001).

We then examine how the semantic content of dream reports changes with dream-level scores of valence and arousal. We compare topic prevalence between dream reports scoring in the top and bottom quartiles for each valence and arousal (Fig. 1B-C). Prevalence denotes the proportion of dream reports in which a topic was present, with 0 indicating that a topic is found in none of the reports and 1 indicating that a topic is present in all of the reports. Qualitatively, heatmaps illustrate distinct profiles of topic prevalence for each affective quartile (Fig. 1B). We next summarize which topics diverged most in prevalence between these strongly affective dream reports. For valence, we find topics related to "abrupt actions" (topic 120), "negation/inaction" (topic 31) and "physical actions" (topic 47) to occur with the greatest prevalence in the most negatively valenced dream reports as compared to the most positively valenced dream reports (Fig. 1C). Action-related topics in negative dream reports may relate to common features of negative dream reports such as running-away, fighting, or the very act of waking-up from a negative dream, whilst the "negation/inaction" topic may relate to the dreamer being unable to perform a desired action (as in sleep paralysis). In comparison, we find that topics related to "human interaction" (topic 95), "dream characters" (topic 20, the most positively valenced dream characters), and "cultural/social context" (topic 9) to occur with the greatest prevalence in the most positively compared to the most negatively valenced reports (Fig. 1C). These findings suggest that highly positive dream reports often take place in collective settings, amongst positive dream characters.

For arousal, we find topics related to "environmental features" (topic 13) and "building materials" (topic 10) with the greatest prevalence in low arousal dream reports compared to high arousal dream reports (Fig. 1C), which suggests that these dream reports emphasize descriptions of the external environment. In contrast, topics denoting "negation/inaction" (topic 31), "abrupt actions" (topic 120), and "shock/concern" (topic 27) have the greatest relative prevalence in high arousal dream reports compared to low arousal dream reports (Fig. 1C). The presence of these themes highlights the overlap between high arousal and negatively valenced dream reports, consistent with the negative correlation between valence and arousal in dream reports.

Additionally, we show the applicability of modern topic modeling towards dream report content analysis by replicating the seminal dream report content analysis performed by Hall and Van de Castle (1966) (39). We compare the DATM-modeled topic content between their original male and female 'norm' dream reports, as the original authors did with a manual coding scheme. Our analysis yields many of the same results, showcasing the power of NLP as a progenitor of manual coding schemes (See Appendix).

Together, these results demonstrate the separability of topics, and thereby narratives, along affective axes (valence/arousal). They also show that the semantic and affective content of dream reports are tightly interrelated.

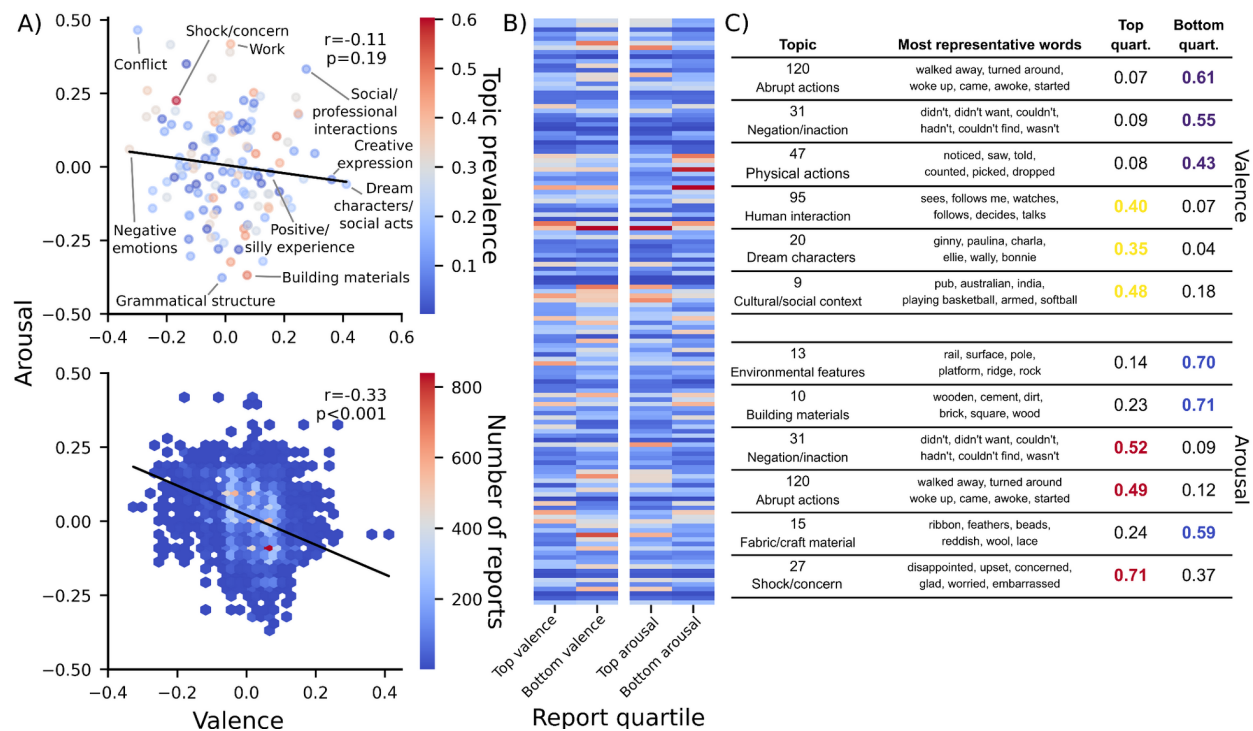

**Figure 1. A:** Top: Scatterplot of valence and arousal for the 130 topics modeled over the DreamBank dataset, coloured by topic prevalence. Prevalence denotes the proportion of narratives in which a topic occurs. More blue points indicate that a topic occurs with lower prevalence and more red, greater. There is no significant correlation between topic valence and arousal (fit: black line; Pearson correlation: r=-0.11[-0.28, -0.06], p=0.19). Bottom: kernel density map of narratives (n=18622) over the dimensions of median dream valence and arousal. The x-axis was divided into 30 hexagonal bins. Dream report valence

and arousal were negatively correlated (fit: black line; Pearson's r=-0.33[-0.34, -0.32], p<0.001). **B:** Heatmaps comparing the prevalence of topics in dream reports in the top and bottom quartiles for valence (left two columns) and for arousal (right two columns). Each horizontal bar represents one topic. **C:** Topic key showing the three topics occurring with greatest divergence in prevalence between top and bottom quartiles for valence (top six rows) or arousal (bottom six rows), ranked by their divergence. The six most representative words for each topic, with the highest cosine similarity to that topic, are shown. Coloured values in bold indicate greater relative prevalence in the corresponding quartile (yellow: top quartile valence; purple: bottom quartile valence; red: top quartile arousal; blue: bottom quartile arousal).

**Networks quantify dream report semantic structure**

Since the affective quality of an experience may also be tied to changes in semantic structure (40, 41), we next quantify the trajectory of dream reports through semantic space. To do this, we create topic networks for each individual dream report. In these networks, each topic is represented by a node and topical transitions are represented by directed edges that connect successive and distinct topics (Fig. 2A). For the following aspect of our analyses, edges were weighted by the cosine similarity between connecting nodes.

First, we characterize these networks in terms of their degree, path length and clustering coefficient, which we measure for each dream report individually. Here, degree refers to the number of unique topics that neighbor a given topic in a particular dream report. Since in and out-degree were perfectly correlated (Pearson correlation: r=1, p<0.001), we focus our analyses upon overall degree. Across all dream report networks, topic degree follows a poisson distribution and the modal topic degree is two (Fig. 2B). Additionally, the average number of steps between any two topics in these networks (average path length) and the longest shortest path (diameter) are lower than in random networks of the same size, whilst the tendency of topics to cluster together (clustering coefficient) is higher than random expectations (mean[95% CI low, 95% CI high], clustering, observed=0.083[0.082, 0.084]; random=0.024[0.022, 0.024]; average path length, observed=3.179[3.168, 3.190]; random=4.673[4.623, 4.724]; diameter, observed=6.971[6.937, 7.004]; random=10.759[10.615,10.903]; effect of randomisation, Linear Mixed-effect Model (LMM), clustering: $\chi^2$=130706, df=1, p<0.001; average path length: $\chi^2$=314789, df=1, p<0.001; diameter: $\chi^2$=8076.70, df=1, p<0.001). These qualities indicate that dream report topic networks have small-world characteristics (42).

Second, we test if the affect of a topic is associated with the affect of its neighbors in a topic network. For each affective dimension (valence/arousal), we found a strong positive association between a topic's affective loading and the average affective loading of its neighbors (Fig. 2B, Pearson correlation between topic affect and mean neighbor affect, valence: r=0.75[0.66, 0.81], p<0.001; arousal: r=0.77[0.68, 0.83], p<0.001). In other words, topics preferentially neighbor topics scoring similarly for valence and arousal within dream reports, indicating that dream reports, in-general, do not experience dramatic shifts in affect.

Third, we quantify the semantic coherence of dream narratives through the mean and standard deviation of signed cosine edge weights in our networks. A more positive mean cosine edge weight indicates that a network predominantly contains edges that connect topics with highly similar semantic content/meanings (i.e. connected topics are close to each other in our embedding), whilst more negative values indicate that a network is composed predominantly of edges that connect topics with more dissimilar semantic content/meanings (i.e. connected topics are far from each other in our embedding). In other NLP work, narratives composed of highly dissimilar words, with more negative cosine similarities to each other, are generally interpreted as being incoherent (43, 44). Incoherence is associated with disorganized thought, as is found in schizophrenia, but also creativity and narrative progression (43, 45, 46). Similarly to in previous work, we also interpret dream report networks composed predominantly of edges that connect unrelated topics, with more negative cosine weights, as incoherent. In our data, topics tend to neighbor topics with more similar than dissimilar meanings (mean coherence=0.26±SD0.20, Fig. 2B), suggesting that on average, dream report networks are composed of coherent semantic connections.

We ask if affect is related to semantic coherence by comparing the mean and standard deviation of semantic coherence (i.e., signed cosine edge-weights) between dream report networks in the top and bottom quartiles for valence and arousal. Our results indicate that both the mean and the standard deviation of semantic coherence are increased in the most positively valenced dream reports compared to the most negatively valenced dream reports; they are likewise increased in the lowest arousal dream reports compared to the highest arousal dream reports (Fig. 2B, Linear mixed-effect Model (LMM), effect of valence quantile on mean coherence: $\chi^2$=76.94, df=1, p<0.001, Cohen's d (d)[95% CI low, 95% CI high]=0.38[0.34, 0.41]); effect of valence quantile on standard deviation of coherence: $\chi^2$=178.19, df=1, p<0.001, d=0.39[0.35, 0.43]; effect of arousal quantile on mean coherence: $\chi^2$=77.77, df=1, p<0.001, d=-1.05[-1.10, -1.01]; effect of arousal quantile on standard deviation of coherence: $\chi^2$=201.04, df=1, p<0.001, d=-0.90[-0.94, -0.86]). In essence, positively valenced and low arousal dream reports are characterized by a mix of semantically coherent topics (high mean) with occasional transitions to distant topics (high standard deviation); the latter events likely reflect narrative transitions to a different domain in semantic space. In contrast, negatively valenced and high arousal dream report networks are composed of consistently incoherent connections, i.e., topics are not typically as related to one another (low mean semantic coherence) but also exhibit few transitions to different regions of semantic space (low standard deviation). Taken together, these results suggest that the affective content of dream reports is associated with structural properties of dream report topic networks.

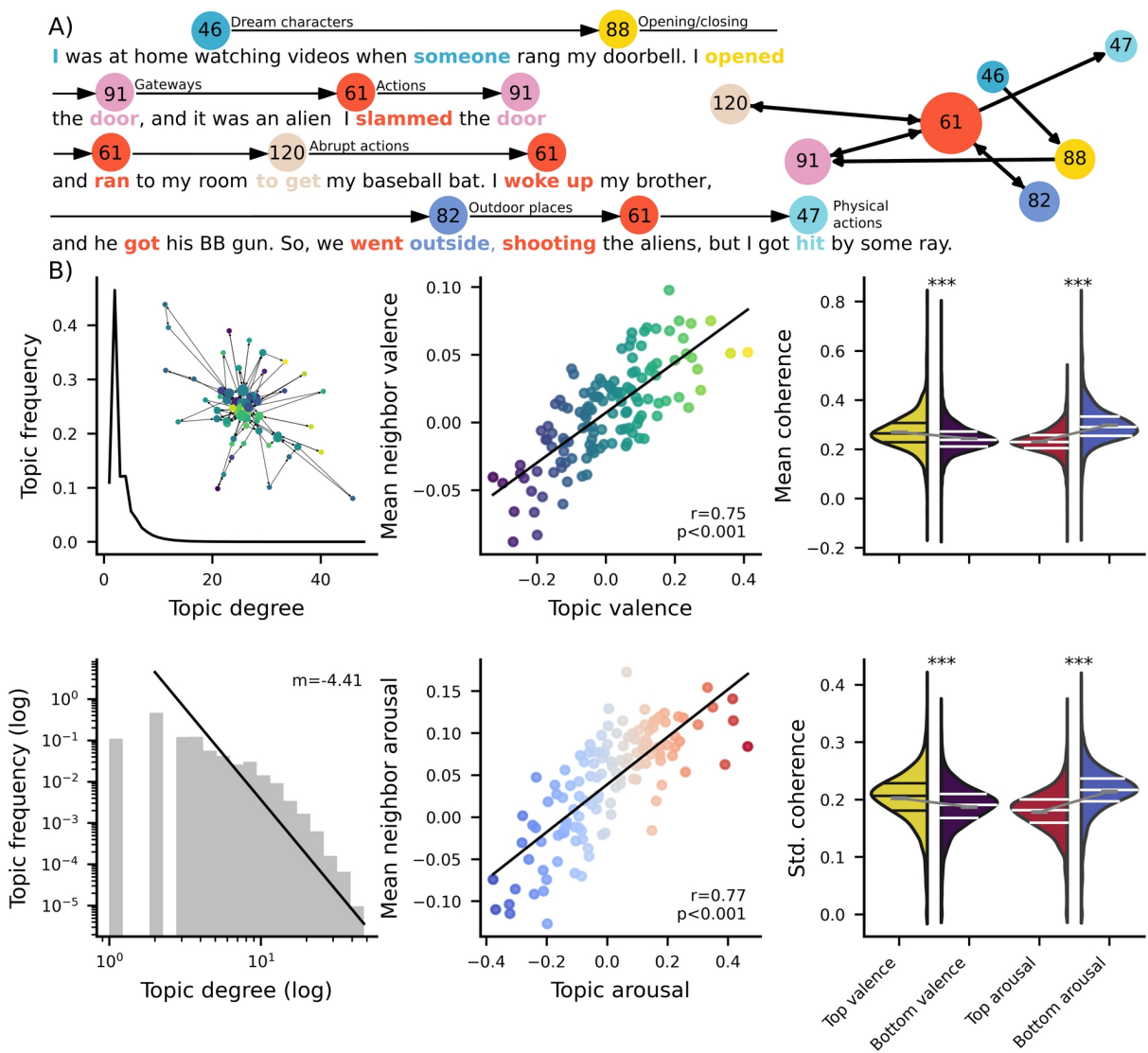

**Figure 2. A:** Example highly negative dream report with topic network. Corresponding topics are placed over their inferred location in the report. The most representative word for each topic is highlighted in bold. Topic numbers and colors correspond to nodes (circles) in the topic network (right), with directed edges (arrows) connecting successive topics. For further examples of annotated dream networks see Appendix. **B:** Characterisation of dream topic networks. Top left: distribution of topic frequency against topic degree (number of connections) over all dream topic networks (n=18622). Insert shows a large topic network, which illustrates the scale-free nature of dream networks (see Appendix for full dream narrative). Nodes are coloured by valence (more positive, more yellow; more negative, more purple) and the size of nodes in networks is proportional to their overall degree. Edge lengths are proportional to the distance between connected topics in our embedding (more semantically similar topics are connected by shorter edges). Bottom left: histogram of log topic degree, divided into 20 bins, against log topic frequency with fit

over topics with degree≥2 (slope=-4.42). Middle column: mean valence of neighboring topics against topic valence (top) or mean arousal of neighboring topics against topic arousal (bottom). Each point represents a unique topic (n=130), coloured by their valence (more negative topics are more purple and more positive, more yellow) or arousal (lower arousal topics are more blue and higher arousal, more red). Mean neighbor valence/arousal was measured by averaging the valence/arousal of each neighboring topic across our entire dataset. Solid lines represent model fits. Topics preferentially attached to topics of similar valence or arousal (Pearson correlation, Valence: r=0.75, p<0.001; arousal: r=0.77, p<0.001). Right column: mean (top) and standard deviation (bottom) of semantic coherence (signed cosine similarity) between neighboring topics in networks scoring in each quartile for valence and arousal. Higher values of cosine similarity indicate neighboring topics to be more semantically similar. Central horizontal black or white lines indicate means and upper and lower lines indicate upper and lower quartiles, respectively. Grey horrizontal bars represent means and 95% confidence intervals (separation too narrow to be visible). Significance of between-quartile LMM comparisons is indicated by asterisks (***: p<0.001).

**Changes in dream report affective content are coupled to topic network structure**

To test if the network structure of dream reports is associated with their affective content, we identify 13 network properties that indicate distinct structural characteristics of narratives (Table 1) and test for their association with either dream report valence or arousal by using partial-least squares regression (PLS). Edge direction and weights are considered, where appropriate. We use three different edge weights for different properties (Table 1, see Appendix: 1) the frequency of directed transitions between connecting topics, 2) the cosine similarity between connected topic vectors, or 3) the inverse of the cosine similarity between connected topic vectors. For measures which use cosine and inverse-cosine edge weights, we ensure that all edge weights are positive by adding 1 to the raw ratings of cosine-similarity, so edge weights can range between zero and two. This ensures that all neighboring topics share a positive relationship and that the strength of the relationship is proportionate to the semantic similarity of the topics. Since negative psychopathologies are known to be associated with reduced linguistic coherence and structure (43, 44), we predict that more positively valenced dream reports will be more structured and linear, whilst more negatively valenced dream reports will be characterized by a disorganized structure, containing more loops and structurally dominant (i.e. high degree) topics.

In agreement with our predictions and as shown in Figure 3A, we find that report valence is positively associated with network properties indicating increased structure (modularity: β=0.0012[0.0004, 0.0020], p<0.001, Cohen's d effect size between top and bottom quartile

(d)=0.04; cosine modularity: β=0.0013[0.0005, 0.002], p<0.001, d=0.04) and linearity (topic path: β=0.0005[-5X10-6, 0.0010], p=0.027, d=0.03; cosine path: β=0.0005[-5X10-6, 0.0010], p=0.027, d=0.03), although the effect sizes were weak. Furthermore, as in our previous analyses (Fig. 2B), we find that the mean and standard deviation of coherence is strongly associated with dream report valence (coherence mean: β=0.0054[0.0047, 0.0060], p<0.001, d=0.38; coherence standard deviation: β=0.0069[0.0047, 0.0060], p<0.001, d=0.39).

In contrast, dream report valence is weakly negatively associated with network properties indicating semantic loops (feedback loops (n.s.): β=-0.0005[-0.0013, 0.0003], p=0.068, d=-0.009; clustering: β=-0.0006[-0.0013, 0.0001], p=0.031, d=-0.05; transitivity: β=-0.0008[-0.0015, -0.0001], p=0.006, d=-0.04); topic number (nodes: β=-0.0011[-0.0017, -0,0006], p<0.001, d=-0.17); disorder and reduced linearity (efficiency: β=-0.0004[-0.0013, 0.0004], p=0.020, d=0.03) but more strongly negatively associated with the presence of dominant topics, as indicated by a degree distribution which is both more variable(degree heterogeneity: β=-0.0018[-0.0024, -0.0012], p<0.001, d=-0.16) and more unequal, whereby few topics share most of the topical connections (Gini: β=-0.0020[-0.0025, -0.0014], p<0.001, d=-0.18); . These network properties explain 10% of the variance in dream report valence (adjusted $R^2$=0.10).

Overall, these results indicate that positively valenced dream reports are structured into distinct but coherent clusters of meaning and follow a linear narrative flow, whilst negatively valenced dream reports contain ruminative loops and highly dominant topics (Fig. 3A).

Next, we test the association between dream report network properties and dream report arousal (Fig. 3B). Our results show that arousal is weakly positively associated with properties indicating topic number (nodes: β=0.0035[0.0024, 0.0045], p<0.001, d=0.09), disorder and topic dominance (degree heterogeneity: β=0.0022[0.0009, 0.0033], p<0.001, d=0.06; Gini: β=0.0020[0.0008, 0.0033], p<0.001, d=0.05), and structure (cosine modularity: β=0.0007[-0.0006, 0.0021, p=0.022, d=-0.016).

In comparison, dream report arousal is found to have weak negative associations with some properties indicating topical loops (transitivity: β=-0.0010[-0.0023, 0.0002], p=0.029, d=-0.04; feedback loops: β=-0.0015[-0.0029, -0.0001], p=0.007, d=-0.04), decreased linearity (efficiency: β=-0.0015[-0.0030, 2X10-5], p<0.001, d=-0.003); and strong negative associations with coherent, but variable, topical transitions (coherence mean: β=-0.0327[-0.0339, -0.0312],

p<0.001, d=-1.05; coherence standard deviation: β=-0.0258[-0.0270, -0.0244], p<0.001, d=-0.90). However, there is no significant association between report arousal and clustering or between arousal and properties indicating overall linearity (clustering: β=0.0005[-0.0008, 0.0026], p=0.190, d=0.004; topic path: β=-0.0001[-0.0012, 0.0008], p=0.711, d=-0.02; cosine path: β=-0.0002[-0.0012, 0.0008], p=0.711, d=-0.02). These network properties explain 31% of the variance in dream report arousal (adjusted $R^2$=0.31).

Taken together, these results indicate that high arousal dream reports are structured around few highly dominant topics. In comparison, low arousal dream reports are characterized by the presence of more topical loops that encircle topics of generally similar meaning, but with occasional transitions to semantically distant themes (Fig. 3A).

We perform further analyses to confirm the robustness of these relationships. In the first of these analyses, we confirm our results were not a consequence of our model hyperparameter selection by rerunning these network analyses over topic models ranging in size from 105 to 255 topics. In the second analysis, we test the robustness of our results to our method of assigning topic valence by, instead, assigning topic valence with the Vader lexicon (47) and rerunning our network analyses. In the third analysis, we test the robustness of our results to our regression method and the sample heterogeneity present in this dataset by performing LMMs of valence/arousal against each network property, with dream source included as a random effect. In the fourth set of analyses, we again test the robustness of our results to the heterogeneity present in the corpus by replicating our network analyses but only on the most homogeneous portion of the corpus, originating from the benchmark Hall and Van de Castle dream report study (39). In the fifth set of analyses, we further examine the source-based variability within the DreamBank corpus by examining each source independently, sub-divided by whether the source was a cohort of dreamers or a longitudinal sample of dream reports from a single subject. In the sixth and final set of analyses, we test if source effects biased our embedding and final results by removing 20 random subsets of dreams from the corpus and, for each subset, retraining the word embedding and topic model and then re-fitting our network analyses. Over each of these analyses, our results were largely replicated, confirming our approach and findings to be highly robust (See Appendix for details).

As an additional control, we next perform a preliminary investigation asking whether the affective-structural relationships measured in dream reports can also be found in waking reports. To do this, we repeat our topic modeling and network analyses on posts collected from the "Diary" subreddit (n=8919), where Reddit users record their daily experiences in a qualitatively similar way to dream reports. We find that the structural properties of waking reports have many different affective associations compared to dream reports. For instance we detected positive associations between diary report valence and topical dominance (degree heterogeneity and Gini) but a negative association between diary report valence and mean coherence (see SI Appendix for further details). Whilst it is challenging to draw strong conclusions from between-corpora comparisons, due to inherent differences in composition and quality, this analysis provides first evidence that the relationships that we detect between dream report affect and structure may not be found in waking reports.

Our previous findings indicate dream report affective content to be associated with semantic content (Fig. 1) and dream report structure (Fig. 3). We next investigate whether affective content is related to the *interplay* between semantic content and network structure. To do this, we measure the relationship between topic prevalence and mean topic degree in dream reports scoring in each top and bottom affective quartile.

Across all dream reports, we find a positive association between topic degree and prevalence (Linear Model (LM), effect of topic prevalence: $F$=602.41, df=1, p<0.001), indicating that the most ubiquitous topics in dream reports are also narratively connected to a greater number of unique themes.

However, we find that the overall affective content of dream reports (i.e. valence and arousal) is associated with significant changes in the dynamic of this relationship. In the most negatively valenced dream reports, the increase in degree with prevalence was significantly greater than in the most positively valenced dream reports (Fig. 3B, LMM, interaction valence quartile×topic prevalence: $\chi^2$=22.59, df=1, p<0.001). Similarly, the increase in topic degree with prevalence was significantly greater in the highest arousal dream reports than the lowest (Fig. 3B, interaction arousal quartile×topic prevalence: $\chi^2$=13.87, df=1, p<0.001).

In further support of this, we find that several of the most prevalent topics in the most negatively valenced and highest arousal dream reports have higher mean degrees than any topic in the most positively valenced and lowest arousal dream reports, respectively (Fig. 3B). This also aligns with our earlier findings from the overall network analyses, which showed that Gini and degree heterogeneity are negatively associated with valence but positively associated with arousal (Fig. 3A). More simply, these findings indicate that the most frequently occuring topics in negatively valenced and high arousal dream reports are even more structurally dominant that the most frequently occuring topics in positively valenced or low arousal dream reports, which may suggest a greater amount of fixation on these dominant themes within the dream.

Taken together, our analyses reveal that the association between the affective and semantic content of dream reports is not only reflected in changes to the prevalence of specific topics but also in the structural role that these topics play within dream networks.

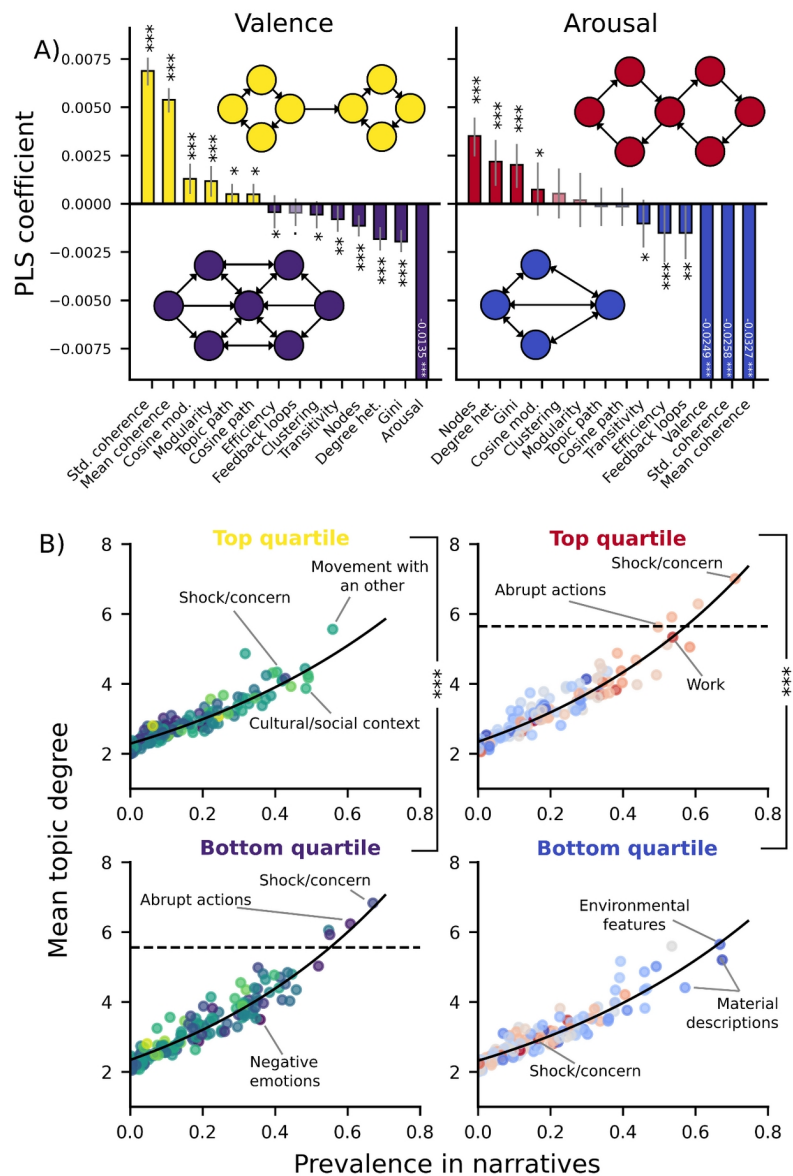

**Figure 3.** Dream report topic network analysis for valence (left column) and arousal (right column). **A:** Partial-least squares regression (PLS) on network properties against valence and arousal over all analyzed topic networks (n=18622). More positive (valence: yellow; arousal: red) PLS coefficients indicate a property to be more positively associated with an affective dimension, whilst more negative (valence: purple; arousal: blue) coefficients indicate a property to be more negatively associated with an affective dimension. Statistical significance of coefficients was obtained via permutation tests (permutations=1000) and is indicated by asterisks (* $p<0.05$, ** $p<0.01$, *** $p<0.001$). Grey whiskers indicate bootstrapped 95% confidence intervals (bootstraps=1000). Inserts of networks illustrate the associated network change. Higher valence topic networks (yellow) are more modular and linear, whilst more negative networks (purple) have more loops (feedback loops (n.s.), clustering and transitivity) and more dominant hubs (degree heterogeneity and Gini). Higher arousal topic networks (red) are larger and have more dominant hubs, whereas lower arousal topic networks (blue) have more loops. **B:** Prevalence of each of the 130 topics in our dataset against their mean degree, averaged over narratives in the top quartile (top row) or the bottom quartile (bottom row) for valence (left) and arousal (right). Solid black lines indicate back-transformed LMM fits. Left: Higher valence topics are more yellow and lower, more purple. Right: Higher arousal topics are more red and lower, more blue. Note the horizontal dashed lines in the graphs for networks of the bottom valence quartile and top arousal quartile. These correspond to the maximum measured mean degree measured for a topic in the datasets of the top quartile of valence and bottom quartile of arousal, respectively.

## Discussion

By combining Discourse Atom Topic Modeling and network analysis, we find that the semantic structure of dream reports is tightly coupled to their affective content. Specifically, our results show that more positively valenced dream reports are more structurally coherent (high mean coherence), structured (high topic/cosine modularity) and more linear (high topic and cosine linearity, low efficiency). By contrast, more negatively valenced dream reports are characterized by semantic loops (high transitivity and clustering) and the dominance of negative topics over the narrative (high Gini and degree heterogeneity; Fig. 3). High arousal dream reports also contained highly dominant topics and incoherent topical connections, but were more structured (high cosine modularity) and contained less semantic loops (low transitivity and feedback loops) than low arousal dream reports. By studying dream reports with the tools of complexity science, our approach offers new avenues for quantitatively understanding the complex interplay between affect, cognition, semantic content, and semantic structure in dream reports and other subjective linguistic reports. Our paper also demonstrates the value of combining nuanced NLP methods (like DATM) with network analysis; indeed, this novel combination extends the power of DATM to shine light on the detailed structure of narratives rather than just their content, as in earlier applications.

Our findings that positively valenced dream reports possess a more linear structure, whilst negatively valenced dream reports contain more thematic loops and structurally dominant topics, support present theories on the role of dreams in emotional regulation. In particular, Kramer posits that a linear organization of thematic content aids successful emotional regulation (high linearity in positive dreams) and that the recurrence of negative themes is an indicator of emotional dysregulation (loops in negative dreams) (20, 21), whilst Hartmann and Nielsen propose that the formation of dense thematic associations towards a highly negative central theme can help one recontextualise and process trauma (high topical dominance in negative dreams) (17–19). Moreover, our results also align with qualitative clinical investigations into the relationship between mood and narrative structure as well as research on emotion and temporal coherence of episodic memories. Several clinical anecdotal studies have linked negative affect with feelings of memory fragmentation (48, 49). Further, studies into the relationship between valence and memory have shown that negative valence is associated with lower memory for surrounding contextual details (50) and a reduced ability to form temporal associations (29, 30), whilst the opposite is true for positive valence (29, 51). Altogether, our findings that positive valence in dream reports is tied to more coherent network structure provides large-scale, quantitative evidence that positive affect may help scaffold narrative coherence for linguistic content.

Our approach to dream report analysis opens up numerous avenues for future research into dream reports and other altered conscious states when presented as natural language. Integrating this method into clinical sleep studies could significantly advance our understanding of dreams by relating dream report content and structure to specific brain activity patterns and other physiological markers. Such an approach could progress not only our understanding of regular dream reports but also of lucid dreaming and disorders of REM sleep. Furthermore, we present preliminary evidence that the affective-structural relationships found in dream reports differ from those in waking reports; future work could build upon this by performing more controlled comparisons. Beyond dream research, our approach for quantifying semantic structure could be developed into a clinical tool to aid in the identification of psychopathologies. It is already known from other clinical research using NLP that conditions such as schizophrenia are marked by reduced coherence between neighboring words in patient speech (44). Network analysis expands upon work in this domain by offering a suite of properties which could be used to profile specific conditions or even indicate specific cognitive processes. For instance, feedback loops, transitivity, and clustering, which indicate topical loops and repetition, could potentially serve as diagnostic markers of rumination, a common feature of many psychiatric disorders (52, 53). However, future work is necessary to ascertain if we can explicitly link network properties to cognitive processes.

Beyond individual symptomatology, our approach could provide insights into how culture shapes the content and structure of experience. In fact, dreams are known to mirror the zeitgeist: a number of studies have already linked the COVID-19 pandemic to an increased prevalence of themes relating to sickness and isolation in dream reports (16). Further, cultural context shapes our waking narratives, and philosophers have long argued that the individual and collective power of emotions is a function of their structural place in a narrative (54, 55). Indeed, contemporary cognitive science has shown that narratives surrounding globally traumatic events can present as aberrant (56) or structured around in-group membership (57). These studies not only support our findings but also, importantly, highlight the diverse applications of our approach beyond dream research towards a richer understanding of how culture interacts with narrative affect, content, and structure to shape subjective experience.

There are several limitations of our study that should be addressed. First, while the DreamBank corpus offers a large and rich dataset of dream reports, the lack of uniform structure of the dataset poses several challenges; namely, the extent to which we are able to control for unknown confounding factors and generalize our findings. While we performed several analyses to confirm the robustness of our results in light of these heterogeneities (see Appendix), future research should collect more uniform datasets with additional ratings of interest (e.g. feelings, sensations, sense of lucidity and agency). Further, it is worth noting that these are retrospective reports of dreams, recorded after the experience of the dream itself. Hence, our approach does not directly analyze dream experiences, but rather the narratives based on episodic memories of these experiences. This introduces two points where our data may be distorted – episodic recall/reconstruction itself and further narrative construction. Indeed, it is known that episodic recall distorts even waking memories (58). Future research should adjust data collection to mitigate some of these issues. For example, the same dream could be probed over time to compare multiple recalls to a baseline representation, or reports could be collected from dreaming subjects in a hypnagogic state.

Finally, in contrast to waking experiences, dream experiences are of situations, happenings, and feelings that are for the most part decoupled from true sensory experience. For this reason, dreams may be closer to waking hallucinatory (59, 60) or imagined experiences (61) than regular waking perception both in terms of processing and phenomenological character (although the debate is still ongoing (62)). This could limit the extent to which the semantic and affective content of dream reports can be accurately modeled, since, relative to waking experience, the character of dreams allows for particular terms or topics to have highly diverse semantic and affective associations, which our model does not account for. For instance, in regular waking experience a book has consistent and predictable sensory and affective characteristics; it has a texture and feel that remains relatively constant throughout experience and it rarely evokes strong affect. In contrast, in a dream, a book can take on any number of sensory characteristics (e.g. it can feel slimy or very hot) and can elicit extreme affect, ranging from joy to terror. Yet, in our model, the "household items" topic, for the term "book" will always have a neutral affective score, even if it occurs in a dream context in which it is evoking terror. Future research should attempt to disentangle the contextual influence on affective content.

In summary, we applied Discourse Atom Topic Modeling and network analysis to reveal that the semantic content, structure and affective content of dream reports are tightly interrelated. Most saliently, we found that network markers of coherent topical connections, linearity, and structure were positively associated with valence in dream reports. These markers also indicated that as the arousal of dream reports increased they became dominated by few topics and contained less coherent topical connections. Remarkably, these findings map onto clinical observations that positively valenced themes can help scaffold narrative and memory formation. This study not only presents a new understanding of the affective and structural interplay of dream reports but also offers a novel methodological approach that can be applied towards studying a diverse range of subjective experiences that have long resisted large-scale scientific analysis, e.g, reports of psychedelic trip experiences, experiences of "pure consciousness" in meditation, near-death experiences, and other altered states of consciousness.

**Acknowledgements and funding sources**
AKB is supported by funding from the National Institute of Health (1DP5OD036172) and a Young Investigator Award from the Brain and Behavior Research Foundation. SR is supported by grants no. 419038924 and no. 397530566 from the Deutsche Forschungsgemeinschaft (DFG, German Research Foundation) as part of the DFG research group Constructing Scenarios of the Past (FOR 2812). JH is supported by the Chateaubriand Fellowship of the Cultural Services of the French Embassy of the United States and by the Foundation for Philosophical Orientation. This project was initiated at the Diverse Intelligences Summer Institute, supported by grant 0333 to JGF from the Templeton World Charity Foundation.

## SI Appendix

### DreamBank

The DreamBank corpus (1) comprises 21341 English-language dream reports collected from various sources and research studies, spanning a diverse demographic of individuals aged 7 to 74 years. This corpus includes reports dating from 1897 to 2011. We scraped the corpus off of DreamBank in July 2023 by using the Python package Beautiful Soup. The mean length of reports in this corpus is 186.86±SD194.38 terms. Dream reports in this database originate both from single individuals, recorded at multiple points in time, and from cohorts of individuals, with each individual being sampled at a single point in time. Since individual-level identities were not always available, we coded each dream's 'source' as either the individual or the group of individuals from which it originated.

To ensure that we trained a high-quality word embedding we applied a number of common pre-processing steps to our data before model training. First, we removed punctuation and converted all text to lowercase. Second, to prevent our model from learning non-dream related text, we removed the interview questions and codings which followed the primary dream narratives of a subset of reports. Third, we applied a bigram transformer to the corpus, which allows commonly adjacent words to be considered as a single term. Fourth, to learn only the most meaning-rich terms, we removed those terms with fewer than 10 occurrences. Last, to prevent the ordering of reports biasing our model, we randomly shuffled the order of reports

We trained our initial embedding over the whole corpus of English language dream reports (n=21341). For learning the DATM, we excluded those extremely short reports containing fewer than 20 terms (2), i.e., two full context windows (n=1418, mean number of terms=12.42±SD5.09). Finally, to prevent our empirical analyses on dream topic content and structure being biased by physical or psychological pathologies, we removed responses from blind participants, a child molester, and traumatized Vietnam veterans (n=470). Our final sample size was 18622 reports.

### Word embedding

We trained word embeddings (3) over the entire DreamBank dataset with Word2Wec (Python package Gensim (4)), using the Continuous-Bag-of-Words architecture and negative sampling (5, 6). We tested 12 embeddings with different combinations of dimensionality (50, 100, 200, 300, 400 and 500) and context window size (4, 8 and 10). Dimensionality refers to the number of dimensions in the semantic space, and thus also of the word vectors therein, whilst context window size refers to the total number of words flanking target words.

We selected our model from amongst these by evaluating their performance with the WordSim-353 Test (7), which compares the model's cosine similarity scores between pairs of words to human-rated similarity scores. Our model, with 100 dimensions and a context window size of 10, scored a Pearson correlation of 0.33, p<0.001 on this test and contained a vocabulary of 12427 terms.

**Topic modeling**

We next used our tuned embedding to train discourse-atom topic models (DATM). In traditional topic models (8–10), topics are modeled as probability distributions over words and documents are modeled as distributions over topics. In DATM, topics remain distributions over words; documents, however, are modeled as trajectories through the semantic space, which can be converted into a sequence of topics. To make the paper self-contained, we give a brief account of DATM here; for details see (2). DATM builds directly on a beautiful theory explaining the property of word embeddings, developed in (11–14).

To identify topics within the semantic space, DATM uses a sparse-dictionary-learning algorithm called K-SVD (K-Singular Value Decomposition). The K-SVD algorithm learns a dictionary of K "atom vectors" such that the word vectors in an embedding can be approximated as a sparse linear combination of these atom vectors. For example, the word vector for "tie" may be a weighted combination of two topics having to do with sports, a topic having to do with clothing, a topic having to do with wires, and a topic having to do with music (3). K-SVD (15, 16) learns these atom vectors by alternately updating the weights in the sparse linear combinations (using orthogonal matching pursuit) and the atom vectors (using singular value decomposition); see the Supporting Information of (2) for a detailed exposition of this step. Each atom vector is converted into a probability distribution over words (i.e., a topic) using a simple log-linear topic model. The probability of a given word being emitted under that topic is proportional to the exponential of the dot product of the atom vector and the corresponding word vector:

$$\Pr(word|atom_i) \propto \exp(\langle \mathbf{atom}_i, \mathbf{w} \rangle)$$

We tested models from 25 to 300 topics. We selected from amongst these models by balancing three quantitative measures of model performance: topic diversity, topic coherence, and $R^2$.

Diversity measures how distinct topics are from one another (17) by calculating the proportion of unique words among the 25 most similar words for each topic in a model. A topic model with a diversity of 1 would have no overlapping words between topics, indicating that the topics are highly distinct. Conversely, a topic model with a diversity of 0 would be composed of topics with identical sets of the 25 most representative words, suggesting redundancy and a lack of semantic breadth. In our study, diversity decreased sharply as the number of topics increased (Fig. S1).

Coherence indicates the interpretability of topics in a model. The measure assesses the semantic similarity of the top 25 most similar word vectors to each topic as their average pairwise cosine similarity (18). Models with high coherence, close to 1, consist of topics represented by highly similar sets of words, indicating that the topics are interpretable and that the representative words for each topic are semantically related. On the other hand, models with low coherence, close to 0, contain topics representing unrelated words, making them less meaningful and harder to interpret. In our analysis, coherence was initially high in models with very few topics before dropping and peaking again, in the model with 105 topics. However, increasing the number of topics beyond 105 resulted in only marginal decreases in coherence (Fig. S1).

$R^2$ measures how well a topic model covers the potential space of meanings in an embedding. More specifically, the metric measures the proportion of variance in the distribution of *word* vectors over an embedding that is explained by the distribution of *topic* vectors over an embedding. A model with an $R^2$ of 1 would perfectly capture the space, indicating that the topics comprehensively represent the underlying meanings within the data. Conversely, a model with an $R^2$ of 0 would cover none of the space, indicating that the topics fail to capture any meaningful semantic information present in the initial embedding. Over our tested models, $R^2$ initially increased rapidly with the number of topics before plateauing after around 105 topics (Fig. S1).

When selecting a model, all three of these criteria are desirable. Based upon these metrics, models with at least 105 topics were indicated to offer high performance (Fig. S1). However, under evaluation of topics for face validity, we found the model with 105 topics contained only a single topic that represented all emotions, both positive and negative. Since emotional granularity was important for our application, we therefore opted to select a model with 130 topics. This model sacrificed only a small amount of diversity and coherence in lieu of more distinct emotional topics. Moreover, the final results of this model were more reliable than for the model with 105 topics when compared to models of different sizes (Fig. S2) and when the valence of topics was evaluated with an external method (Fig. S3). It should also be noted that our results were consistent even when we used larger, and more emotionally granular, models than our selected 130-topic model (Fig. S2). Our final model obtained scores of 0.87 for diversity, 0.54 for coherence and 0.79 for $R^2$. We assigned labels to the topics of our final model based upon their top-25 most representative words.

With our selected model, we next assigned sequences of topics to each narrative in the corpus. DATM accomplishes this by mapping each document into a trajectory through semantic space and then discretizing that trajectory into a sequence of topics. We first move a 10-term sliding window through each narrative one term at a time, capturing the local semantic context around each term as a context vector. We then obtained an embedding for the corresponding context vector by using the Smooth Inverse Frequency (SIF) embedding technique (11, 12). SIF first creates a context vector by calculating the weighted average of the word vectors within the corresponding context window; a vector's weight is determined by the inverse frequency of the term across the full corpus, thereby reducing the influence of common but less informative words. Next, a global context vector is created by sampling a range of context windows over the full corpus, computing the corresponding context vectors, and then computing the first principal component of these. The global context vector is, therefore, the direction that explains the most variance in contextual meanings across the whole corpus. To extract just the meaning of the local context window, SIF takes the weighted context vector and sutracts off its projection onto the global context vector, to obtain a SIF embedding. Finally, we can then assign a topic to a context window, at its corresponding position in the narrative, by measuring the topic vector with the highest cosine similarity to the SIF embedding. As the window progresses through the narrative a sequential string of topics is generated, providing a dynamic representation of the thematic elements within the narrative. For full details see (2).

**Valence and arousal assignment**

As explained in the main text, we assigned topics ratings of valence and arousal by loading topics onto independent dimensions for each emotional axis. We used seed words to represent each affective pole in our embedding. We created a valence dimension by first creating positive and negative valence vectors from six seed words each. To create the valence vectors we average the embeddings for the six seed words. Seed words for the negative valence vector were: bad, sad, scared, anxious, angry and disgust. 'Bad' corresponds to generally negative experience, whilst the remaining seed words denote the most commonly occurring negative emotions in dreams and encompass the major Western negative emotional categories (19–22). Seed words for the positive valence vector were polar to the negative seed words: good, happy, safe, confident, relaxed and love. We then created a valence dimension by subtracting the negative valence vector from the positive.

Similarly, we created an arousal dimension by subtracting a low arousal vector from a high arousal vector. Seed words for arousal were selected to balance the valence for each high and low vector by including three seed words associated with positive valence and three associated with negative valence (22). The high arousal seed words were: energetic, delighted, excited, tense, angry, and frustrated; and the low arousal seed words were: peaceful, relaxed, calm, lazy, depressed, and tired.

Next, we computed the valence and arousal of topics, within the context of dreams, by calculating their cosine similarity to these dimensions. Topics with the most positive valence would have a cosine similarity of 1, whilst topics with the most negative valence would have a cosine similarity of -1. Neutral topics would have a cosine similarity of 0. Finally, we obtained scores of report-level valence and arousal by measuring the median affective loading of all topics within a dream report.

**Network creation and analysis**

We analyzed the semantic structure of dream reports by creating dream topic networks. These networks represent topics as nodes and the transitions between successive and distinct topic vectors, $i$ and $j$, as directed edges. There were no self loops in these networks. In other words, an edge could not lead into the same topic from which it projected. We used three separate edge weights, $\omega$, for different network properties.

1) the frequency of transitions, $f$, between topics in a narrative:

$$\omega_{ij} = f_{ij}$$

2) The cosine similarity between connecting topic vectors (which we interpret as semantic coherence):

$$\omega_{ij} = cosine(i, j) + 1$$

3) The inverse of the cosine similarity between connecting topic vectors:

$$\omega_{ij} = \frac{1}{cosine(i,j)+1}$$

Frequency edge weights, which increase with the number of transitions between topics, were used to calculate topic modularity. Non-inverse cosine edge weights, which increase with more closely shared meaning between two topics, were used to calculate cosine modularity. Networks with inverse cosine edge weights can be thought of as spatial networks over an embedding; their edge weights increase with distance in meaning between topics and so were used to calculate cosine path. Topic path, diameter, average path, and efficiency, were measured over undirected, unweighted networks. degree heterogeneity, Gini, feedback loops, clustering and transitivity were measured over directed, unweighted networks.

Network properties were analyzed using *Networkx* 3.0.1 for Python 3.8 (23). For modularity, we defined community partitions over the network by using the Louvain algorithm. We used Dijkstra's algorithm to solve topic and cosine paths. The mean number of nodes, or unique topics, in our networks was 27.70±SD17.79.

We created randomized dream networks by performing an edge swapping procedure. In this process, we randomly selected pairs of edges in the original network and swapped their endpoints. This edge swapping procedure was repeated 2000 times, for each network, to ensure that the resulting randomized networks had a significantly different topology from the original network while maintaining the same number of nodes and edges and preserving the degree distribution of the original network. To characterize dream networks, we then compared average clustering, diameter, and average path length between observed and randomized dream networks by using LMMs considering the effect of randomisation and by contrasting means and 95% confidence intervals.

**Validation**

To ensure that our results were not a chance result arising from our choice of topic model size, we tested the reliability of our final network analyses across models containing 105, 130, 155, 205 and 255 topics. To do this, we followed the same general analysis procedure for each of these models: first assigning topics, second loading topics onto dimensions of valence and arousal, third computing median topic affective loadings over each report, fourth extracting topic networks from each report, fifth taking network properties from these networks and, finally, computing PLS coefficients for network properties, with either valence or arousal as a response. We then compared our final PLS coefficients between topic models of different sizes for each of valence and arousal (Fig. S2).

We found our results to be highly consistent between models of different sizes. For valence, coefficients for all network properties, except efficiency and cosine edge weight (coherence) standard deviation, consistently trended in the same direction (Fig. S2). Efficiency was positively associated with valence only in the model with 205 topics, whilst cosine edge weight (coherence) standard deviation was negatively associated with valence only in the largest model, with 255 topics. Amongst all other models, PLS coefficients were consistently negative for efficiency and consistently positive for cosine edge weight standard deviation (Fig. S2).

For arousal, we found model size to have no influence on the sign of PLS coefficients for transitivity, cosine edge weight mean, cosine edge weight standard deviation, Gini, degree heterogeneity and node (distinct topic) number. In models with 130 topics and above, efficiency and feedback loops were always negatively associated with arousal, whilst modularity and cosine modularity were always positively associated with arousal. Topic path and cosine path were negatively associated with arousal in four out of five of our tested models (105, 130, 155 and 255 topics, but not 205 topics). Last, clustering was positively associated with arousal in three out of five of our tested models (105, 130, and 205 topics, but not 155 and 255 topics) (Fig. S2).

We were next interested in testing the reliability of our network analyses against an external method for assigning valence to topics and reports. We did so by using the Vader sentiment lexicon (24) to score topic sentiment. Vader is a widely used dictionary of English words with associated human-annotated sentiment scores. The lexicon does not provide scores for arousal, so we limited this validation to valence only. We used the lexicon to rate sentiment scores of the top 25 most representative words for each topic, weighted by word cosine similarity to the topic, before summing the scores to obtain topic-level sentiment scores. It should be noted that many of the terms in our embedding were not present in the Lexicon, so we could not score the sentiment for many of our topics with this method (n=15); this limitation of Vader further highlights the power of our new approach. Nevertheless, cosine valence ratings and Vader sentiment ratings were correlated amongst those topics which Vader could rate (Model with 130 topics: Pearson correlation=0.34, p=0.0002). After rating topic-level valence with this method, we next replicated our entire network analysis pipeline. Again, we took median topic-level sentiment scores over each Vader-rated report to obtain report-level scores, which were used as a response variable in PLS models, taking dream network properties as predictors. We then tested the correlation of these coefficients against those obtained using our cosine ratings of valence, as in the main text (Fig. 3A), for topic models trained on 105, 130, 155, 205, and 255 topics. We found positive correlations between the PLS coefficients obtained using Vader and valence ratings obtained via cosine similarity in models with 130 topics and up (Fig. S3, Pearson correlation for model with 130 topics: r=0.74, p=0.004). However, the sign of all coefficients agreed between both methods only for our selected model, with 130 topics (Fig. S3).

We took two approaches to confirm the robustness of our results to the heterogeneities inherent in the DreamBank corpus. First, we performed univariate linear mixed-effects models (LMMs) on each of our network properties against report valence and report arousal. These LMMs accounted for source-based variation by including dream source as a random effect. LMM results are shown alongside corresponding PLS results in Table S1. Interestingly, LMMs found no significant association between node number and either valence or arousal (valence: p=0.061; arousal: p=0.927). This finding further confirms that our results for the other network properties of interest were not a consequence of the scaling of properties with network size.

We now compare PLS and LMM results for valence and arousal. For valence, the sign of the PLS and LMM-generated estimates agrees for all properties; although LMM p-values were non-significant (but still<0.10) for clustering, topic path and cosine path. For arousal, the sign of LMM-generated estimates agreed with PLS-generated estimates for all statistically significant properties except Gini and degree heterogeneity; LMMs found non-significant negative associations between these properties and arousal. Although disagreement between LMMs and permutation tests may indicate that the association between Gini and report arousal may result from source effects, it may also be a product of correlations between Gini and other network properties, which univariate LMMs do not account for.

Therefore, we further confirmed the robustness of our empirical analysis by performing a more limited analysis, restricted to the homogenous Hall and Van de Castle (HVdC) male and female 'normal' dreams (25) (n=797). These dreams were collected from a diverse cohort of participants and, alongside the HVdC dream content coding scheme, continue to be used as a baseline for studies investigating dreams (26–28). The HVdC research was seminal in the field of dream science and was one of the first comparative studies of dream content.In their study, HVdC compared dream content between males and females. HVdC made several key findings in this comparison: female dreams contain more emotional content, more social interactions, and more female dream characters than male dreams, whilst male dreams contain more physical aggression and are more often spent in outdoor settings than female dreams. In recent years, several studies have attempted to replicate the HVdC study with more modern NLP approaches (29, 30). However, these approaches are distinct from DATM (and other approaches classically defined as topic modelling, such as LDA). More specifically, these studies extracted topics by computing the frequency of different word classifications (e.g. the word "parent" would be classified into the "family" class) in the HVdC dataset and then corresponded these word classifications to the preexisting HVdC coding scheme (29, 30). These approaches focus purely upon the lexical composition of texts, do not encode the contextual meanings of words within a text, and are constrained to only identifying "topics" that were also identified in the original HVdC schema or in the human-coded word classification. In contrast, DATM starts with word embeddings which, by taking sliding context-windows through a text, allows a consideration of the contextual meanings of words within a text. Then, by leveraging K-SVD, DATM can meaningfully discover and encode the semantics of different topics (as embeddings), as they occur within the specific context of dreams. Critically, the topics created by DATM are derived from an unsupervised clustering approach; these topics represent the discursive building blocks of a particular corpus and are not constrained to a manually-developed coding scheme. We use DATM to replicate the original HVdC study on dream content by comparing the prevalence of topics in male and female dreams from the HVdC dataset. Our quantitative DATM content analysis agreed with the original HVdC analysis (Fig. S4). We found the three topics with greatest relative prevalence in female dreams, as compared to male dreams, corresponded to an emotional category, "shock/concern" (topic 27, female prevalence=0.60, male prevalence=0.43); "dream characters" (topic 46, female prevalence=0.35, male prevalence=0.23); and "relationships" (topic 22, female prevalence=0.35, male prevalence=0.24). Additionally, we found topics relating to "meetings/conversations" (topic 38, female prevalence=0.32, male prevalence=0.22)," venues/gathings" (topic 3, female prevalence=0.24, male prevalence=0.14) and "female" (topic 6, female prevalence=0.10, male prevalence=0.06) to occur with greater relative prevalence in female dreams as compared to male dreams. In contrast, the four topics with the greatest relative prevalence in male dreams corresponded to "outside environmental features" (topic 13, female prevalence=0.24, male prevalence=0.32); "movement with an other", which includes aggressive actions such as "push" (topic 104, female prevalence=0.11, male prevalence=0.17); "outside movement" (topic 81, female prevalence=0.18, male prevalence=0.23); and "tools", such as "rifles" and "bows" (topic 100, female prevalence=0.06, male prevalence=0.11).

Next, we repeated our network analyses on the HVdC data and measured the correlation between the PLS network results performed on the HVdC data to those from the full corpus. Despite comprising less than 5% of the entire dataset, PLS coefficients from our HVdC analysis were highly correlated with those from our full dataset (Fig. S5, Pearson correlation between full corpus and HVdC PLS coefficients of network properties against valence: r=0.86, p<0.0001; Pearson correlation of full corpus and HVdC PLS coefficients between network properties against arousal: r=0.95, p<0.0001). For the HVdC PLS of network properties against valence, only coefficients for node number and feedback loops opposed the sign of those coefficients from our full analysis; for arousal, only clustering and cosine/topic path differed in sign between our full and HVdC analysis (Fig. S5).

Furthermore, since the HVdC is gender-labeled (female, n=436; male, n=360), we took advantage of the dataset to test for the influence of any gender biases in our results by performing separate models on HVdC male and female datasets. PLS coefficients were highly correlated between genders for each affective dimension (Fig. S5. Pearson correlation, Valence coefficients: r=0.93, p<0.001; Arousal coefficients: r=0.94, p<0.001). Furthermore, with the exception of Gini and transitivity, the sign of PLS coefficients for all properties agreed between male and female data. Gini and degree heterogeneity were found to be positively associated with valence in females, but not males, and Gini was negatively associated with arousal in females, but not males. Transitivity and node number were positively associated with arousal in females, but negatively associated with arousal in males (Fig. S5).

We now test for inter-source variability in the relationships between affect and structure by performing our network analyses on each source independently. We consider the cohort and longitudinal sources in DreamBank separately. For each source, we compare the correlation of the PLS coefficients over dream report network properties as measured from the full dataset and the proportion of properties that share the same coefficient sign. Strong agreement between a source-level analysis and our original analysis is indicated by more positive correlation coefficients and a larger proportion of properties sharing the same sign. Additionally, to test if certain dream report network properties are more or less robust to source effects, we also test, for each network property, the proportion of sources for which the same sign was measured as in the full dataset. Importantly, we do not necessarily predict that exactly the same associations measured over the full corpus will also be measured in each subset, since it is expected that individuals and cohorts differ from one another in the content and flow of dream narratives. For cohort sources (Fig. S6), we find that PLS coefficients are strongly correlated with those measured over the full dataset for both valence and arousal (mean Pearson correlation coefficient±SD, valence=0.61±0.24; arousal=0.78±0.16). Furthermore, we find that, on average, most of the signs of the PLS coefficients measured for each cohort source agreed with those measured over the full dataset (mean proportion of sign agreement±SD, valence=0.71±0.15; arousal=0.66±0.17). Interestingly, the cohort source with the lowest agreement (miami-lab) is a cohort of participants engaging in a laboratory sleep study—which is not the case for the other sources—suggesting that the setting in which individuals either dream or report their dreams could influence the structure of dream reports. For dream report properties, we find that the sign of their measured associations with valence and arousal, on average, agrees with the full dataset, when measured separately on each cohort source (mean proportion of sign agreement±SD, valence=0.71±0.15; arousal=0.66±0.22). For valence, node number is the only property which is measured to agree across cohort subsamples less than 62% of the time (agreement=0.46). For arousal, clustering, cosine modularity, feedback loops, modularity, and node number were measured to agree across cohort subsamples less than 50% of the time. However, the properties which we measured to have stronger effects in the main manuscript (Gini, degree heterogeneity, measures of path length, mean coherence, standard deviation of coherence, and arousal) agreed across cohort subsamples ≥69% of the time.

For longitudinal sources (Fig. S7), we measure slightly weaker, but still relatively strong, correlation coefficients across PLS coefficients taken from each subsample (mean Pearson correlation coefficient±SD, valence=0.47±0.31; arousal=0.68±0.24). As for the cohort analysis we find that, on average, there is relatively high sign agreement when measured both across sources  (longitudinal source-level mean proportion of sign agreement±SD, valence=0.65±0.215; arousal=0.62±0.13) and when measured across properties (property-level mean proportion of sign agreement±SD, valence=0.75±0.15; arousal=0.62±0.19). Interestingly, since certain individuals are represented by multiple sources (e.g. from different points in their life), we can observe that the affective-structural associations measured appear highly similar across their different samples (Fig. S7, samples: ["b", "b-baseline", "b2"]; ["bea1", "bea2"]; ["jasmine1", jasmine2", "jasmine3", "jasmine4"]; ["madeline1-hs", "madeline2-dorms", "madeline3-offcampus", "madeline4-postgrad"]; ["phil1", "phil2", phil3"]), supporting the notion that a basis for inter-source variability in narrative structure is a consequence of individual variability in dreamt experience or dream recounting.

Last, we test if our results could have been a consequence of heterogeneities in the DreamBank corpus biasing the initial word embedding. To do this, we use an established approach for testing bias in embeddings (29) that is highly-similar to -fold cross-validation, which is a proven methodology for evaluating the validity of machine learning models (30). Specifically, we randomly removed five percent of the dream reports from the DreamBank corpus and iterated this procedure, without replacement, until each dream report had been removed from the corpus once. This provided a total of 20 subsets of the DreamBank corpus, each comprising 95% of the dream reports in the original corpus. Then, for each sub-corpora we retrained the word embedding and topic model using exactly the same parameters as for our full data. We then fit the topic model originating from each subcorpus to the entirety of the dream reports from the original corpus and subsequently performed the network analyses and PLS procedure. Again, we would not expect to measure identical affective structural associations in each of these subset analyses, since the analysis is performing on a reduced semantic and affective space; we, therefore, expect the model to capture less semantic and emotional granularity. The results for this validation are shown in Fig. S8. Surprisingly, these results show that the relationship between dream report valence and arousal can be somewhat variable, which is presumably a consequence of the aforementioned effects of corpus-size reduction. However, we find that even when valence and arousal become decoupled, most of the structural relationships measured with our original model are highly robust (mean Pearson correlation coefficient±SD, valence=0.40±0.26; arousal=0.96±0.0.03). As we find with the full corpus, clustering, degree heterogeneity, Gini, feedback loops, and transitivity were all found to have negative associations with dream report valence for every subset analysis, whilst mean coherence, cosine/topic path, and modularity were found to have positive associations with dream report valence ≥65% of the time. Furthermore, we find all network properties, except clustering and cosine modularity, to share the same directional associations with dream report arousal ≥85% of the time. We do find that the standard deviation of coherence was positively associated with dream report valence only 25% of the time, but maintained its association with dream report arousal 100% of the time, which indicates that this structural feature could be more tied to report arousal than it is to report valence. These results indicate that the heterogeneous composition of the DreamBank corpus exerts relatively little bias on our measured associations between affect and structure.

Combined, these analyses demonstrate that our results are highly robust to our choice of model, method, and data. Additionally, we invite readers to examine the face validity of our findings for themselves through example networks (Fig. S6-7) and by exploring our attached Jupyter Notebook and data (see Supplementary Material), which allows users to sample dreams from the corpus alongside information on their valence/arousal, topic content, and network properties, as well as visualize their networks. The output for this is illustrated in Supplementary Figures 10 and 11.

Comparison to waking reports

To test whether the relationships between affect and structure that we detected are unique to dreams or apply also to waking experience we next analysed a set of waking reports, from the "Diary" subreddit. In this subreddit, users journal their everyday experience in a way that can be considered qualitatively similar to a recounting of a dreamt experience. All posts in this subreddit through to December 2024 were downloaded via the Pushshift reddit dump released on July 2024. The corpus comprises 8919 reports (mean length=247.79±284.68SD). Prior to training any models on this dataset, we first preprocessed the dataset so as to remove any images, links, html tags, and emojis — such that we retained only the textual data of the diary entries. We then performed the exact same general methodological procedure on this dataset: we trained a word embedding (dimensions: 100; window size: 10; WordSim-353 Pearson correlation=0.20, p<0.001), trained a topic model (130 topics; Coherence=0.53, Diversity=0.69; $R^2$=0.82), computed topic and narrative-level valence and arousal, computed structural graphs and associated network properties for each narrative, and performed PLS to test the association between affective content and semantic structure. Our analyses reveal the Diary valence and arousal at the report-level are positively associated with one another (Fig. S9, association of arousal with valence: β=0.0190[0.0158, 0.0204]; association of valence with arousal β=0.0249[0.0215, 0.0278], d=0.31) and that there are altogether different associations between valence and structural properties of narratives in the Diary reports, as compared to in dream reports (Fig. S9). For instance, we find that Diary report valence is positively associated with properties indicating topical dominance (Fig. S9, degree heterogeneity: β=0.0024[0.0007, 0.0036], p<0.001, d=0.04; Gini: β=0.0012[-0.0005, 0.0025], p<0.001, d=0.02), dense connectivity (Fig. S9, efficiency: β=0.0012[-0.0010, 0.0033], p<0.001, d=0.02; clustering:

β=0.0011[-0.0006, 0.0024], p=0.014, d=0.02), and overall topic number (Fig. S9, nodes: β=0.0018[0.0004, 0.0028], p<0.001, d=0.03). In contrast, valence is shown to be negatively associated with properties indicating increased structure (Fig. S9, cosine modularity: β=-0.0012[-0.0030, 0.0007], p<0.001, d=-0.02; modularity: β=-0.0015[-0.0032, 0.0006], p<0.001, d=-0.02) and coherent, but variable, topical transitions (Fig. S9, mean coherence: β=-0.0046[-0.0060, -0.0027], p<0.001, d=-0.08; standard deviation of coherence: β=-0.0010[-0.0028, 0.0010], p=0.0153, d=-0.02). Furthermore, we find no significant associations between diary report linearity and valence (Fig. S9, topic path: β=0.0007[-0.0005, 0.0019], p=0.052, d=0.01; cosine path: β=0.0007[-0.0005, 0.0019], p=0.052, d=0.01). However, consistent with our main analyses, we do find that valence is weakly negatively associated with the presence of topical loops (Fig. S9, transitivity: β=-0.0011[-0.0026, 0.0002], p=0.0140, d=-0.02; feedback loops: β=-0.0010[-0.0029, 0.0007], p=0.023, d=-0.02). Although several associations were statistically significant (p<0.05), we note that for a majority of properties the effect sizes were very weak and that for number of properties, the 95% confidence intervals overlapped with zero (Gini, efficiency, clustering, cosine modularity, and standard deviation of coherence). This suggests that while certain associations are statistically significant by conventional thresholds, the magnitude and direction of these effects should be interpreted cautiously, given the uncertainty around the estimates.

We find that a number of structural properties share the same associations with arousal in diary reports as in dream reports, although the effect sizes are generally very weak. First, we find that the arousal of diary reports similarly shares positive associations with properties indicating increased topical dominance (Fig. S9, degree heterogeneity: β=0.0091[0.0069, 0.0103], p<0.001, d=0.12; Gini: β=0.0051[0.0030, 0.0064], p<0.001, d=0.06), overall topic number (Fig. S9, Nodes: β=0.0038[0.0021, 0.0050], p<0.001, d=0.05), and increased clustering (Fig. S9, clustering: β=0.0030[0.0009, 0.0048], p<0.001, d=0.04). In addition, we find that diary report arousal is similarly negatively associated with coherent, but variable topical transitions (Fig. S9, mean coherence: β=−0.0174[−0.0199, −0.0142], p<0.001, d=-0.23; standard deviation of coherence: β=−0.0109[−0.0133, −0.0078], p<0.001, d=-0.14). In contrast to what was measured in dream reports, we find the arousal of diary reports is weakly positively associated with the number of reciprocal connections in a narrative (Fig. S9, feedback loops: β=0.0034[0.0013, 0.0056], p<0.001, d=0.04) and linearity (Fig. S9, topic path: β=0.0016[0.0000, 0.0032], p=0.002, d=0.02; cosine path: β=0.0016[0.0000, 0.0032], p=0.001, d=0.02), but negatively associated with properties indicating increased structure (Fig. S9, modularity: β=−0.0029[−0.0051,

−0.0003], p<0.001, d=-0.04; cosine modularity: β=−0.0024[−0.0047, 0.0002], p<0.001, d=-0.03).
Last, we find no significant associations between diary report arousal and either efficiency or
transitivity (Fig. S9, efficiency: β=0.0006[−0.0022, 0.0033], p=0.147, d=0.008; transitivity:
β=−0.0009[−0.0030, 0.0009], p=0.078, d=-0.01).

Overall, our analyses of these diary reports provides a first indication that, whilst their are some
commonalities,  dreamt and waking reports may bear different affective-structural associations.
However, we stress that the presented comparison of dreamt and waking reports is preliminary
and, due to differences in data quality and collection methods, an accurate comparison of the
affective-structural associations in dreamt versus waking reports is not permitted between these
two corpora. Future work should seek to collect more carefully curated corpora that combine
dreamt and waking experiences.

**Statistics**

To investigate the effect of randomisation on network properties; the associations between
dream emotions and network properties; and the relationship between affect, mean topic
degree, and topic prevalence, we used Linear Mixed Effect Models (LMMs) (R 3.4.1, R
package: lme4 (31)). In all LMMs, except those testing for the relationship between mean topic
degree, prevalence, and affect, dream source was included as a random effect. In cases where
we compare the top and bottom affective quartiles, "quartile" was coded as a categorical
variable. In all other cases valence and arousal were coded as continuous variables. All network
properties were coded as continuous variables. For LMMs testing the effect of network
randomisation, either clustering, average path, or diameter were included as a response;
"randomisation" was included as a categorical predictor (observed or random); and the index of
the original dream was included as an additional random effect (each index corresponds to one
dream, with an observed and randomized network). For LMMs testing the association between
mean topic degree, prevalence, and affect, mean topic degree was included as a response and
the interaction between topic prevalence and affective quartile was included as a predictor (Fig.
3). For LMMs investigating the association between affect and network properties (Table S2),
valence and arousal were taken as response variables, and network properties were taken as
predictors. In these LMMs, topic identity was included as an additional random effect. We
ensured that LMM assumptions were satisfied by testing the normality of residuals with
histograms and Q-Q plots. The significance of LMMs was tested by type 2 ANOVA with Wald's
test.

For each structural measure we use Cohen's D to provide a standardised measure of effect size of the property in the top quartile for valence or arousal. Cohen's D is calculated as the mean of the property in the top quartile minus the mean of the property in the bottom quartile divided by the pooled standard deviation. The absolute value of Cohen's D corresponds to the strength of the effect (0.2≈weak effect; 0.5≈medium effect; 0.8≈strong effect) and the sign of the effect corresponds to a positive (+) or negative (-) association with affect (32).

Since network properties are often highly correlated, we incorporated all properties into a partial-least square regression (PLS) (Python package scikit-Learn (33)). PLS handles correlations between predictors by reducing the predictors to a selected number of latent variables, which maximize the covariance between the predictors and response. Absolute values of PLS coefficients correspond to the relative effect of a predictor on a response. Coefficients with a positive sign indicate a property to be positively associated with a response, whilst negative coefficients indicate a property to be negatively associated with a response. For PLS models taking report valence as a response we accounted for correlations of network properties with report arousal by including it as an additional predictor. Similarly, for PLS models taking report arousal as a response, we included report valence as a predictor. Before performing each PLS, we accounted for differences in property scale by rescaling predictors between -1 and 1, centered on 0.

In PLS the number of latent variables into which the predictors are reduced is determined by the number of components included in the model. We determined the optimal number of components by evaluating the performance of PLS models with 1-10 components. For each of these models trained over the full corpus and full HVdC data, performance was evaluated by measuring $R^2$ in k-fold cross-validation ($k$=5). Here, k-fold cross validation splits the dataset into five equal-sized parts, trains the PLS model on four out of five of these parts and tests the model on the fifth part. This process was then iterated five times, so that each part of the data was used in both training and testing. At each iteration of the cross-validation, the model's performance was assessed by $R^2$, which measures the proportion of variance in the test data that is explained by the model trained on the training data. Finally, we obtained an overall $R^2$ score by averaging $R^2$ over the five iterations. For each of our models trained on the full data, $R^2$ increased marginally with additional components beyond 1. These PLS obtained an overall $R^2$ of 0.10 for valence and 0.31 for arousal. For the full HVdC data, we selected a model with two components for valence ($R^2$=0.040) and a model with one component for arousal ($R^2$=0.22). Due to the small size of the separate HVdC male and female data, we ensured that there was enough data for model training by performing leave-one-out cross validation. This method is similar to k-fold cross validation but leaves only a single data point out of the training process and trains the model on the remaining data. This is then repeated for all data points. For this method, we quantify model performance as the relative decrease in Root Mean Squared Error (RMSE) of our model fitted over this data from a baseline model, which predicts the mean of the target variable for all instances. As we did for the previous models, this is then averaged over all iterations of validation. Since RMSE measures the difference between the observed and model-predicted values (as the standard deviation of model residuals), a decrease in this metric is desirable for a model. For both male and female HVdC data, we selected models with one component for both valence and arousal, with which which we measured improvements in RMSE (relative decrease in RMSE (%), valence, male: 2.08; valence, female: 11.52; arousal, male: 2.34; arousal, female: 12.91).

We used two complementary approaches to assess the certainty of generated PLS coefficients. First, to test the statistical significance of the PLS coefficients for our optimized model, we implemented a permutation test. At each permutation, we disrupted any relationships between the response and predictors by randomly shuffling the values of the response across rows of our data. Next, we tested for the same relationship between network properties and valence/arousal by refitting the PLS on the permuted data, which provides coefficients corresponding to the relationship between our original predictors and a random response. We obtained a distribution of permuted coefficients by iterating this process 1000 times. Next, we obtained empirical two-sided p-values for the association between each predictor and valence/arousal by computing the proportion of times the absolute value of each original PLS coefficient exceeded the absolute values of the permuted coefficients. Last, we protected against multiple testing by applying Benjamini-Hochberg correction to these p-values.

Second, to indicate the plausible range of the true coefficients we extracted 95% confidence intervals around PLS coefficients. To generate confidence intervals we iteratively extracted 1000 bootstrap samples from our data and refit the model on each resampled data. Bootstrap samples were equal in size to the original fitted data but included random rows, such that certain rows of data might be repeated, whilst others might be omitted. Across the 1000 bootstrapped PLS-coefficients we then calculated 95% confidence intervals as the 2.5th and 97.5th percentiles, for each coefficient.

Together, these two approaches provide both a direct hypothesis test, to reflect the probability of observing an effect as extreme as ours under the null hypothesis that dream report network structure and affect share no associations (permutation testing), and an estimate of effect size variability and reliability (bootstrapped confidence intervals).

**Example dream topic networks**

To help improve the familiarity of readers with our approach, we now provide a series of highly negative and highly positive dreams from the DreamBank dataset, which each provide illustrative examples of topic network properties. First, we report examples of highly negative dreams. Corresponding topic networks are illustrated in Figure S6, alongside the percentile position of associated dream network properties. Narratives are shown alongside the associated topic sequence and topic labels. These dreams have limited linearity and show topical loops and high degree hubs.

**Negative dream 1:**

"I came over early to my boyfriend's house to surprise him. I walked back into his bedroom and found him lying in bed naked with some other girl. I can't quite remember who it was (maybe his ex-girlfriend.) This dream made me feel very upset. I woke up and began to cry and then fell back to sleep."

Topic sequence:
'51: Daily events', '3: Venues and gatherings', '120: Abrupt actions', '111: Returning', '12: Rooms', '111: Returning', '120: Abrupt actions', '55: Body postures', '12: Rooms', '55: Body postures', '42: Personal attributes and relationships', '27: Shock/concern', '7: Experience and perception', '121: Negative emotions', '120: Abrupt actions'

**Negative dream 2:**

"I dreamt about my deceased paternal grandpa Ernest. He was there at some family gathering. Then, he was dying. Everybody else just pretty much stood around. I think my father said something to him, and I knelt down and gave him a hug and said goodbye...I was happy I got to say goodbye, but I was upset that nobody else seemed to care- it didn't seem right that someone should go and their family doesn't seem to care. Then I remember something of crawling into a little cubbyhole."

Topic sequence:
'46: Dream characters 2', '86: Dream characters 3', '46: Dream characters 2', '27: Shock/concern', '9: Cultural/social context', '27: Shock/concern', '120: Abrupt actions', '41: Questions and indefinite references', '120: Abrupt actions', '125: Body parts', '32: Loving acts', '27: Shock/concern', '103: Seeming', '27: Shock/concern', '103: Seeming', '27: Shock/concern', '103: Seeming', '85: Intention', '103: Seeming', '34: Modal verbs', '38: Couple acts', '1: Time periods', '58: Dream recall'

**Negative dream 3:**
"There was a young woman with a very spirit-like quality to her. She wasn't frightening really, but could be described as evoking some sense of anxiety. There was also another person present, but I can't identify whether it was a man or a woman. The second person kept trying to approach the first, as they got near, the spirit-like woman would suddenly disappear and reappear a further distance away. This happened several times. Throughout, one of them, I think the spirit-like woman, kept calling, Tarkington. Tarkington. I think it was the name of the other person. It was somewhat dark and I believe they were outside. The last time the spirit-like woman moved she seemed to float through the wall of a building. The building was one of many along a street, side-by-side with no space between them, like you would see in a city where the business buildings are all connected. But the buildings seem to be much smaller than normal - not a size that an actual person could walk into, yet bigger than what I would describe as miniatures. The person who had been trying to approach the woman went up to the building that she seemed to float into and was peering inside, wishing they could go in."

Topic sequence:
'26: Age and gender categories', '11: Descriptions of experience', '121: Negative emotions', '31: Negation/inaction', '57: Comfort', '44: State and condition', '11: Descriptions of experience', '7: Experience and perception', '74: Work', '26: Age and gender categories', '83: Social/professional interactions', '26: Age and gender categories', '83: Social/professional interactions', '41: Questions and indefinite references', '26: Age and gender categories', '74: Work', '1: Time periods', '111: Returning', '91: Gateways', '13: Environmental features', '8: Nature', '77: Time and movement', '7: Experience and perception', '26: Age and gender categories', '58: Dream recall',

'74: Work', '26: Age and gender categories', '74: Work', '26: Age and gender categories', '97: Travel problems', '27: Shock/concern', '7: Experience and perception', '113: Weather', '7: Experience and perception', '1: Time periods', '7: Experience and perception', '111: Returning', '7: Experience and perception', '103: Seeming', '111: Returning', '103: Seeming', '105: Building features', '10: Building materials', '107: City locations', '119: Quantities', '98: Transportation', '85: Intention', "112: Communicating 'nothing'", '107: City locations', '17: Intention', '107: City locations', '57: Comfort', '11: Descriptions of experience', '44: State and condition', '85: Intention', '7: Experience and perception', '85: Intention', '57: Comfort', '7: Experience and perception', '74: Work', '103: Seeming', '26: Age and gender categories', '36: Movement and direction', '89: Trying', '91: Gateways', '103: Seeming', '107: City locations', '103: Seeming', '85: Intention', '88: Opening/closing'

Next, we report examples of highly positive dreams. Corresponding topic networks are illustrated in Figure S7. Narratives are shown alongside the associated topic sequence and topic labels. Percentile position of associated dream network properties in the corpus are shown. These dreams are highly linear; dream 2 is also highly modular.

**Positive dream 1:**
"I heard or saw 1 and four. I was working. There was a lot of brass work. John Sorenso, the boss, was making a fireplace ornament out of a brass bar. He twisted it into a fancy shape."

Topic sequence:
'47: Physical actions', '119: Quantities', '17: Intention', '9: Cultural/social context', '35: Actions and interactions', '15: Fabric/craft material', '10: Building materials'

**Positive dream 2:**
"Something about working as a go-between for a fat operator type delivery warehouse trucker. The scene is on the corner of Albany and Summerset St. in back of the Murray's house, the street I grew up on. This guy is dropping boxes and bundles all over the truck and I'm a little kid. I say something to the effect that I've been watching out for opportunities for him and report a couple. That is opportunities I guess for some kind of a wheeler dealer operation."

Topic sequence:
'25: Speculation and association', '9: Cultural/social context', '15: Fabric/craft material', '9:

Cultural/social context', '105: Building features', '107: City locations', '105: Building features', '82: Outdoor places (urban)', '63: Household items', '100: Tools and weapons', '122: Take-out food', '63: Household items', '98: Transportation', '96: Talking', '74: Work', '96: Talking', '1: Time periods', '76: Personal attributes', '1: Time periods', '119: Quantities', '18: Categories and comparisons'

**Positive dream 3:**
"Riding in a car with Zena's boyfriend; Wally is in the back seat eating a chocolate doughnut; He gives us a bite; He looks good and healthy; Happy."

Topic sequence:
'40: Housing locations', '22: Relationships', '40: Housing locations', '22: Relationships', '9: Cultural/social context', '35: Actions and interactions', '60: Ordering food and drink', '95: Human interaction', '32: Loving acts'

**Supplementary Figures and Tables**

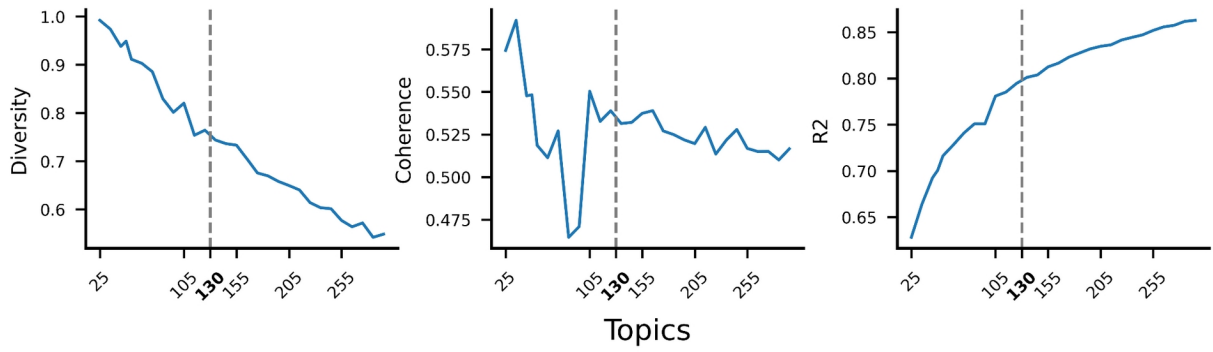

**Figure S1.** Measures of topic diversity, coherence and $R^2$ for DATM topic models ranging in size from 15 to 305 topics, which we used to determine our final model size. Our selected model, with 130 topics, is highlighted in bold.

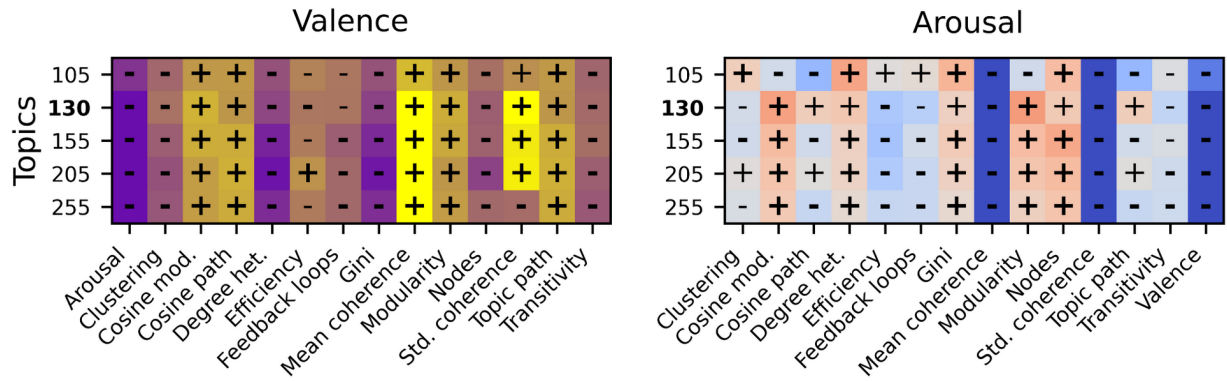

**Figure S2.** Matrices of final PLS coefficients for the effect of our tested network properties on valence (left) and arousal (right) at different topic model sizes. - indicates a negative PLS coefficient (valence: more purple; arousal: more blue) and + indicates a positive coefficient (valence: more yellow; arousal: more red). +/- operators in bold indicate significant coefficients (p<0.05), as determined by a permutation test.

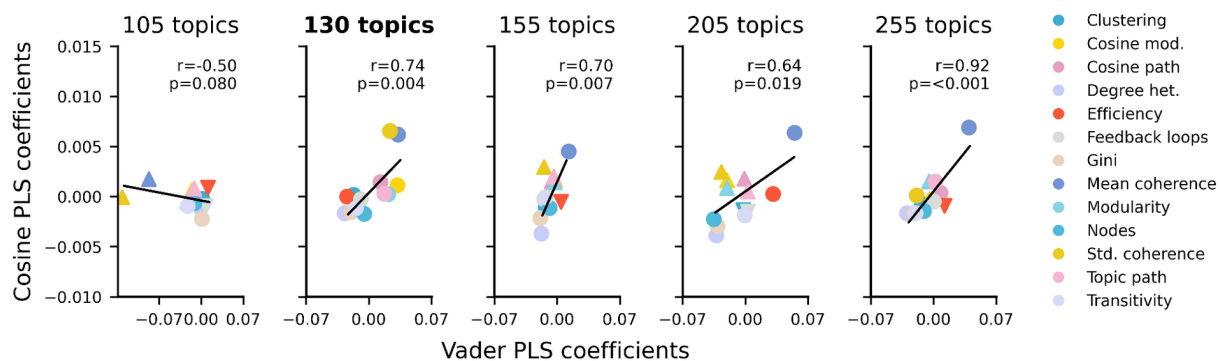

**Figure S3.** Correlation between partial least squares (PLS) coefficients for the effect of network properties on average report topic valence, where topic valence is computed either by topic cosine similarity to a valence dimension (y-axis) or by topic sentiment ratings from the Vader lexicon (x-axis). For the Vader lexicon method, the cosine-weighted lexicon ratings of the top 25 most representative words for each topic were summed to rate topic valence. Pearson correlation coefficients (r) with p-values are given in the top right corner of each plot. Each point represents a network property. Upward triangles indicate instances where the sign of the cosine PLS coefficient was positive and the Vader coefficient was negative, while downward triangles indicate a negative cosine PLS coefficient and a positive Vader coefficient. Circles indicate agreement in PLS coefficient sign between both methods of rating topic valence. Note the strong correlation and agreement in coefficient sign for our final model containing 130 topics (bold).

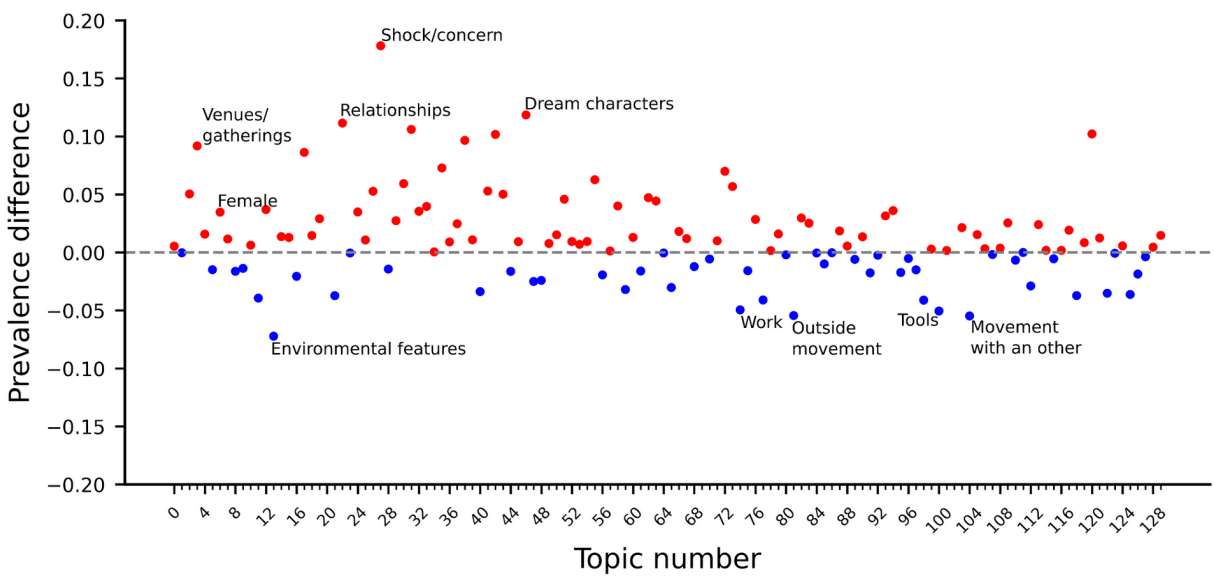

**Figure S4.** Gendered comparison of topic prevalence on the HVdC data. For each topic, the prevalence in male narratives was subtracted from the prevalence in female narratives. More female topics are more positive (red), while more male topics are more negative (blue).

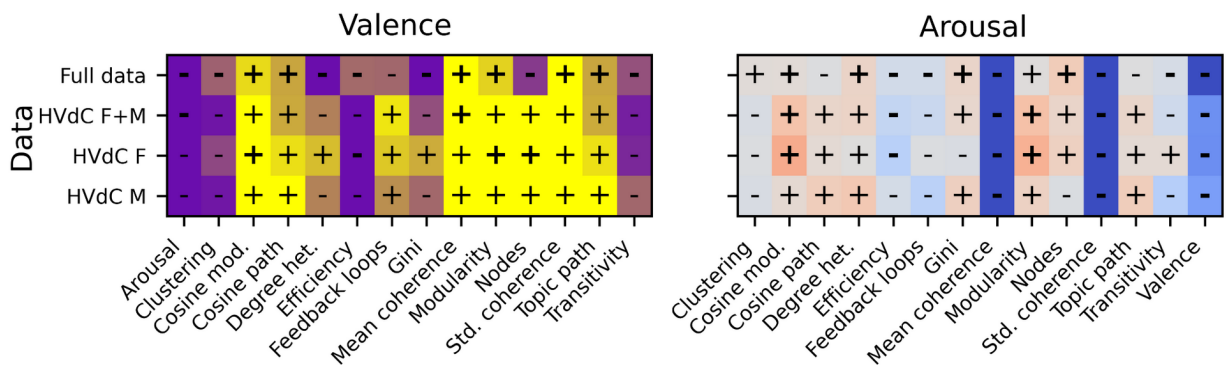

**Figure S5.** Matrices of PLS coefficients from our full dataset against those from the full and gender-separated HVdC data. - indicates a negative PLS coefficient (valence: more purple; arousal: more blue) and + indicates a positive coefficient (valence: more yellow; arousal: more red). +/- operators in bold indicate significant coefficients (p<0.05), as determined by a permutation test.

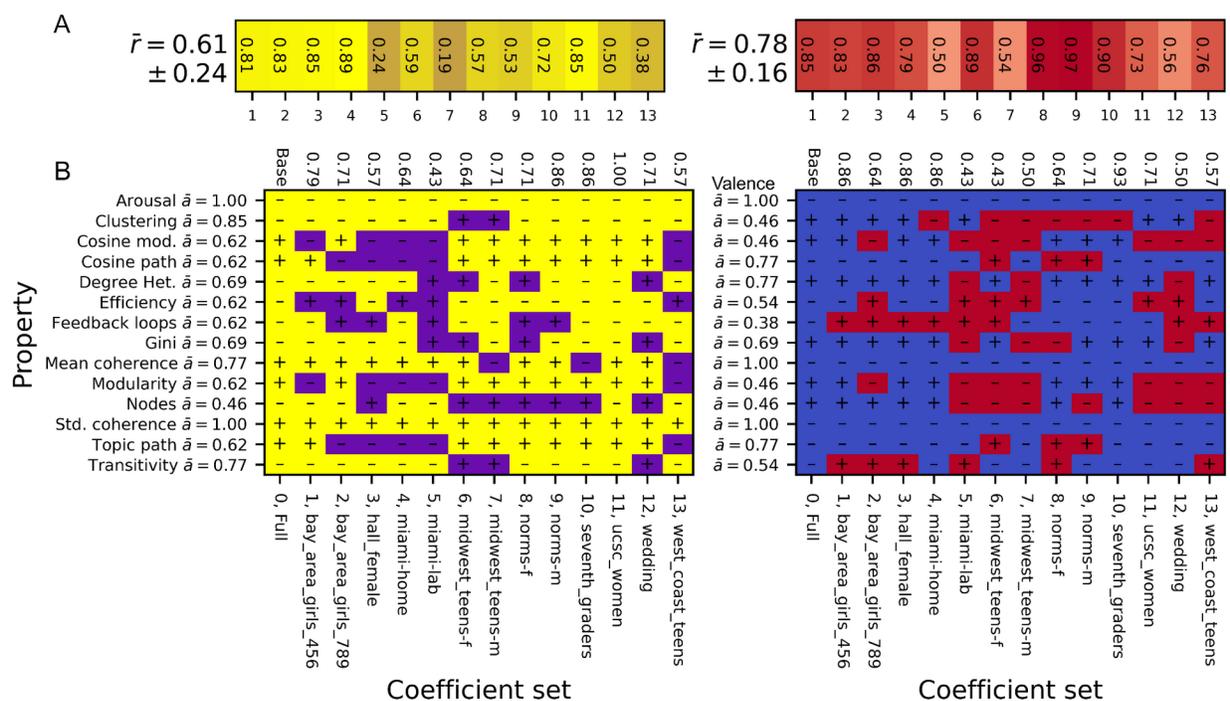

**Figure S6.** Robustness analysis comparing the results of dream report network analyses performed on the whole dataset to each cohort source within the DreamBank dataset. **A:** Pearson correlation coefficients between the PLS coefficients taken over the full DreamBank dataset and each cohort source for valence (left) and arousal (right). More positive correlation coefficients (valence: more yellow; arousal: more red) indicate closer agreement with the results taken over the full corpus. The average correlation coefficient ($\bar{r}$ ±SD), taken across all subsets is provided. **B:** Matrices displaying agreement in coefficient sign with the full dataset (first column) for valence (left) and arousal (right). Each column represents a source in DreamBank and each row represents a property. + indicates a positive PLS coefficient, whilst - indicates a negative coefficient. Cells are coloured by their agreement with the full dataset (valence: yellow=agreement, purple=disagreement; arousal: blue=agreement; red=disagreement). The proportion of cells that agree with the full dataset ($q$) is shown for each property as row labels and for each source at the top of each column.

**Figure S7.** Robustness analysis comparing the results of dream report network analyses performed on the whole dataset to each longitudinal source within the DreamBank dataset. **A:** Pearson correlation coefficients between the PLS coefficients taken over the full DreamBank dataset and each longitudinal source for valence (top) and arousal (bottom). More positive correlation coefficients (valence: more yellow; arousal: more red) indicate closer agreement with the results taken over the full corpus. The average correlation coefficient ($\bar{r} \pm$SD), taken across all subsets is provided. **B:** Matrices displaying agreement in coefficient sign with the full dataset (first column) for valence (top) and arousal (bottom). Each column represents a source in DreamBank and each row represents a property. + indicates a positive PLS coefficient, whilst - indicates a negative coefficient. Cells are coloured by their agreement with the full dataset (valence: yellow=agreement, purple=disagreement; arousal: blue=agreement;

red=disagreement). The proportion of cells that agree with the full dataset ($\underline{a}$) is shown for each property as row labels and for each source at the top of each column.

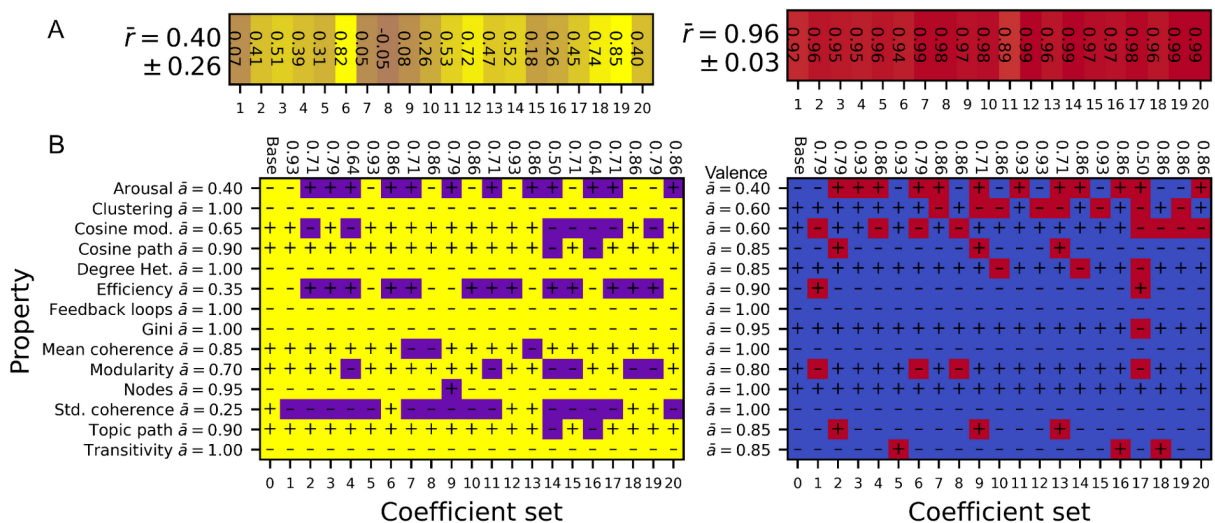

**Figure S8.** Robustness analysis testing for the influence of bias in word embeddings on network analyses. 20 random subsets of entries equal in size to 5% of the corpus were removed and, for each removal, the word embedding and topic model was retrained. For each retrained topic model, we refit the topics to the full DreamBank dataset, which were then used to construct dream report networks. **A:** Pearson correlation coefficients between the PLS coefficients taken over the full DreamBank dataset and each retrained model for valence (left) and arousal (right). More positive correlation coefficients (valence: more yellow; arousal: more red) indicate closer agreement with the results measured with the embedding trained over the full corpus. The average correlation coefficient ($\underline{r} \pm$SD), taken across all fitted models is provided. **B:** Matrices displaying agreement in coefficient sign with the embedding trained over the full dataset full dataset (first column) for valence (left) and arousal (right). Each column represents one of the 20 retrained word embeddings and each row represents a property. + indicates a positive PLS coefficient, whilst - indicates a negative coefficient. Cells are coloured by their agreement with the full dataset (valence: yellow=agreement, purple=disagreement; arousal: blue=agreement; red=disagreement). The proportion of cells that agree with the full dataset ($\underline{a}$) is shown for each property as row labels and for each retrained model at the top of each column.

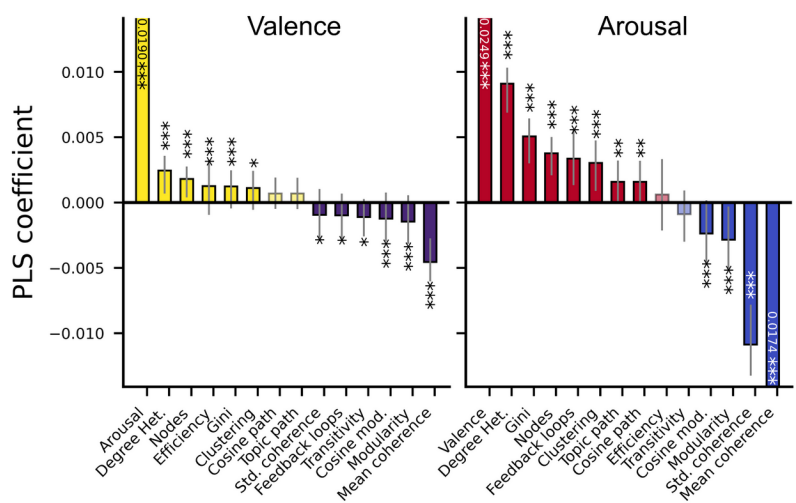

**Figure S9.** Diary report topic network analysis for valence (left column) and arousal (right column). **A:** Partial-least squares regression (PLS) on network properties against valence and arousal over all analyzed topic networks (n=8919). More positive (valence: yellow; arousal: red) PLS coefficients indicate a property to be more positively associated with an affective dimension, whilst more negative (valence: purple; arousal: blue) coefficients indicate a property to be more negatively associated with an affective dimension. Statistical significance of coefficients was obtained via permutation tests (permutations=1000) and is indicated by asterisks (* p<0.05, ** p<0.01, *** p<0.001). Grey whiskers indicate bootstrapped 95% confidence intervals (bootstraps=1000).

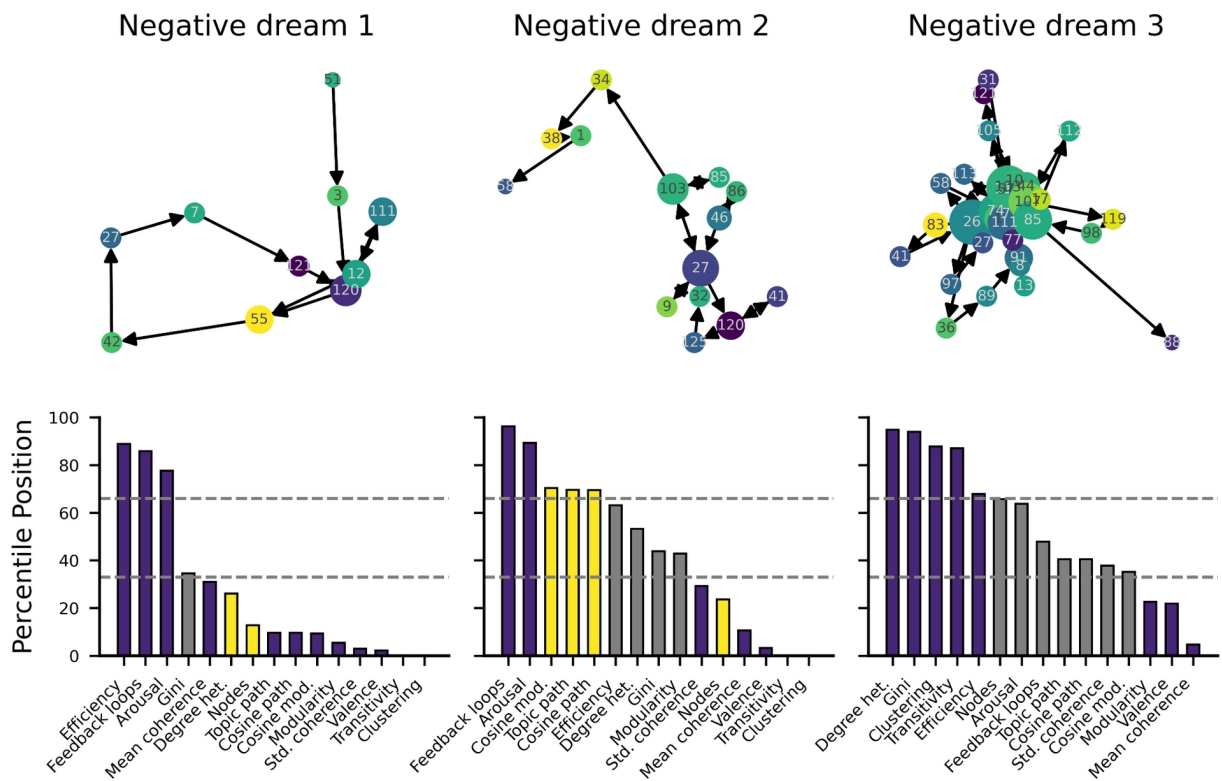

**Figure S10.** Top networks: nodes (topics) are coloured by valence (more yellow: more positively valenced; more purple: more negatively valenced). The size of nodes is proportional to their overall degree (number of connections). Directed edges connect towards successive topics in a narrative. Edge lengths are proportional to the inverse of the cosine distance between connected topics (topics closer together in meaning are connected by shorter edges). Bottom row: barplots illustrating the percentile position of dream network properties. Properties that are associated with negatively valenced dreams (i.e. the property ranks in the bottom 33% for that property in our corpus and is positively associated with valence or the property ranks in the top 33% for that property in our corpus and is negatively associated with valence) are colored in purple, whilst properties that are associated with positively valenced dreams (i.e. the property ranks in the bottom 33% for that property in our corpus and is negatively associated with valence or the property ranks in the top 33% for that property in our corpus and is positively associated with valence) are colored in yellow. Properties that fall in the middle 33% are colored in gray. Upper and lower gray dashed lines represent the 67th and 33rd percentiles, respectively. For negative dream 3 (as shown in Fig. 2B), note the particularly high degree heterogeneity and Gini, as would be expected for a scale-free network.

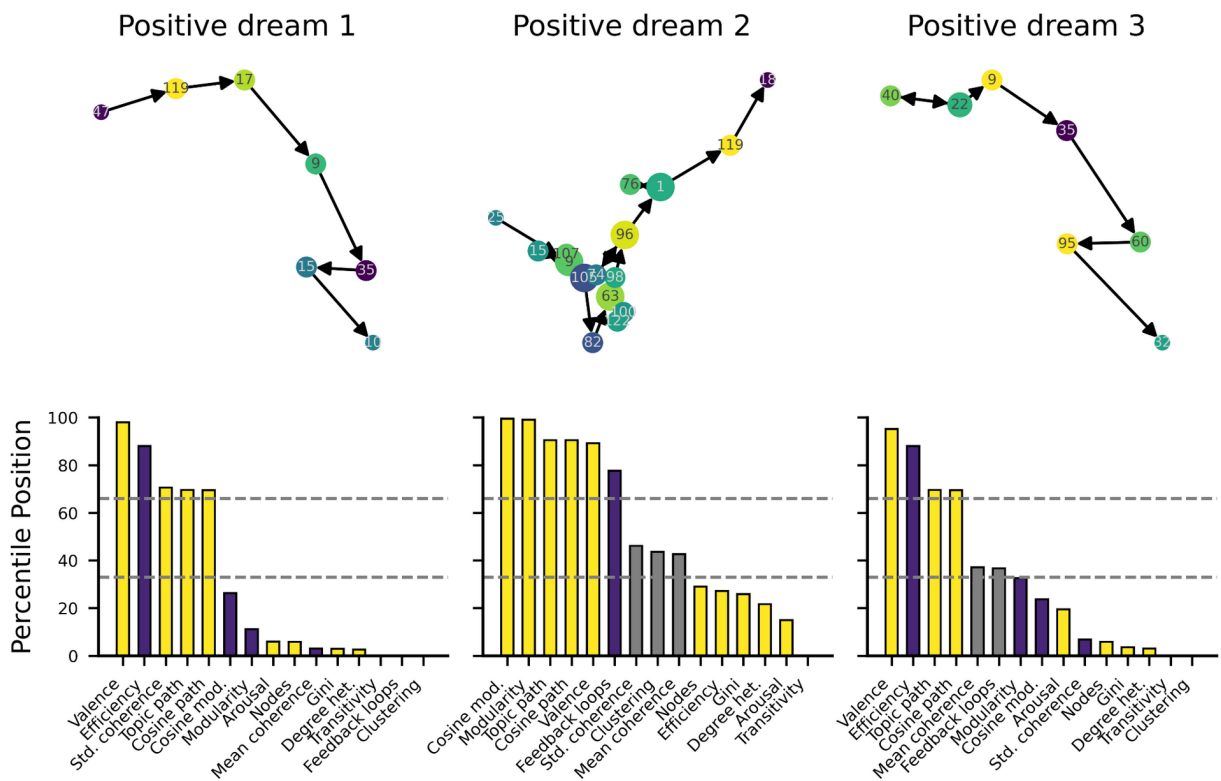

**Figure S11.** Example highly positive dream networks. Top networks: nodes (topics) are coloured by valence (more yellow: more positively valenced; more purple: more negatively valenced). The size of nodes is proportional to their overall degree (number of connections). Directed edges connect towards successive topics in a narrative. Edge lengths are proportional to the inverse of the cosine distance between connected topics (topics closer together in meaning are connected by shorter edges). Bottom row: barplots illustrating the percentile position of dream network properties. Properties that are associated with negatively valenced dreams (i.e. the property ranks in the bottom 33% for that property in our corpus and is positively associated with valence or the property ranks in the top 33% for that property in our corpus and is negatively associated with valence) are colored in purple, whilst properties that are associated with positively valenced dreams (i.e. the property ranks in the bottom 33% for that property in our corpus and is negatively associated with valence or the property ranks in the top 33% for that property in our corpus and is positively associated with valence) are colored in yellow. Properties that fall in the middle 33% are colored in gray. Upper and lower gray dashed lines represent the 67th and 33rd percentiles, respectively. Note the particularly high modularity of positive dream 2, reflecting the high degree of compartmentalisation in this network.

**Table S1.** Topic from our topic model with 130 topics. The ten most representative words and the valence and the arousal of each topic are shown.

| Topic no. | Topic label | Most representative terms | Valence | Arousal |
|---|---|---|---|---|
| 0 | Conversation | said_oh, replied, implying, said, knows, says, says_oh, kept_saying, replies, saying | 0.009 | 0.174 |
| 1 | Time periods | week, month, few_days, hours, weekend, weeks, year, half_hour, afternoon, an_hour | 0.096 | 0.252 |
| 2 | Wonderment | asked, wondered, ask, kept_asking, asking, wasnt_sure, dont_know, wonder, wonders, asks | 0.077 | 0.065 |
| 3 | Venues and gatherings | school, camp, party, church, university, meeting, hotel, any_rate, college, restaurant | 0.016 | 0.390 |
| 4 | Body parts | heart, ear, attitude, breast, shoulder, knee, shoulders, butt, throat, finger | -0.017 | 0.175 |
| 5 | Interpersonal dynamics | convince, trust, resist, cheat, impress, listen, allow, ignore, kill, assure | 0.086 | 0.125 |
| 6 | Female | she, her, herself, mary, shes, ellie, charla, her_cunnilingus, shed, baby | -0.156 | 0.032 |
| 7 | Experience and perception | seen, done, discovered, known, forgotten, notion_that, begun, been, previously, eaten | -0.036 | 0.152 |
| 8 | Nature | ocean, sky, river, sea, waves, clouds, storm, lake, waters, earth | -0.056 | -0.106 |
| 9 | Cultural/social context | pub, australian, india, playing_basketball, armed, softball, eating_lunch, friends_nat, armstrong, jr | 0.146 | -0.135 |
| 10 | Building materials | wooden, cement, dirt, brick, square, wood, structure, metal, concrete, curved | 0.074 | -0.369 |
| 11 | Descriptions of experience | distinct, apparent, humor, highly, incredibly, sophisticated, impressive, vivid, overall, pleasurable | -0.053 | -0.195 |
| 12 | Rooms | room, bedroom, apartment, kitchen, bathroom, living_room, office, closet, hotel_room, classroom | -0.057 | 0.097 |
| 13 | Environmental features | rail, surface, pole, platform, ridge, rock, pavement, wire, wheels, swings | 0.018 | -0.324 |
| 14 | Auxiliary verbs | must_have, might_have, had, ive, weve, hadnt, ive_never, havent, have, hasnt | -0.035 | 0.051 |
| 15 | Fabric/craft material | ribbon, feathers, beads, reddish, wool, lace, ink, silk, cotton, fur | 0.063 | -0.210 |
| 16 | Directional prepositions | toward, towards, across, against, onto, away_from, along, above, forward, behind | -0.006 | -0.259 |
| 17 | Intention | supposed, going, here, where, what, okay, sure, getting_ready, gon_na, why | 0.212 | 0.084 |
| 18 | Categories and comparisons | some_sort, some_kind, kind, great_deal, sort, instead, type, all_kinds, all_sorts, reminded_me | -0.146 | -0.030 |
| 19 | Remembering | remember, cant_remember, recall, dont_remember, dont_recall, happening, cant_recall, bizarre, dont_know, figuring_out | 0.020 | -0.018 |
| 20 | Dream characters 1 | ginny, paulina, charla, ellie, wally, bonnie, ramona, annie, dwight, dovre | 0.213 | -0.110 |
| 21 | Motion | rolled, pushed, bent, slid, pulled, floated, slipped, laid, broke, jumped | -0.125 | -0.164 |
| 22 | Relationships | best_friend, brother, uncle, partner, sister, girlfriend, cousin, roommate, aunt, | 0.072 | 0.070 |

| | | grandmother | | |
|---|---|---|---|---|
| 23 | Spatial descriptions | in, into, filled_with, Ishaped, womens, remodeled, walking_through, dimly_lit, sharing, enter | -0.063 | -0.212 |
| 24 | Capabilities | happen, make, die, indicate, use, have_been, do, fit, understand, be_able | 0.175 | 0.104 |
| 25 | Speculation and association | about, perhaps, related, maybe, responsible_for, suspected, general, involved, discussing, associated_with | 0.032 | 0.124 |
| 26 | Age and gender categories | young_man, woman, man, boy, girl, young_woman, kid, priest, lady, guy | -0.043 | 0.173 |
| 27 | Shock/concern | disappointed, upset, concerned, glad, worried, embarrassed, very_upset, pleased, sick, surprised | -0.167 | 0.225 |
| 28 | Past events and states | was, had_been, saw, remembered, ended_up, noticed, kept, had_gotten, worked, fell_asleep | -0.208 | 0.120 |
| 29 | Transfer | bring, come, take, give, brought, send, sent, call, go, pick | 0.087 | 0.154 |
| 30 | Times of day | morning, late, early, since, night, day, next_day, evening, until, friday | -0.098 | 0.006 |
| 31 | Negation/inaction | didnt, didnt_want, couldnt, hadnt, couldnt_find, wasnt, didnt_know, wanted, wouldnt, knew | -0.237 | 0.193 |
| 32 | Loving acts | smile, lovingly, sadly, smiles, love, kiss, kisses, laugh, hug, say_yes | 0.055 | -0.009 |
| 33 | Categories of people | women, students, men, other_people, children, guys, girls, kids, babies, employees | 0.119 | 0.240 |
| 34 | Modal verbs | ill, must, will, can, wont, should, cant, theyll, shall, might | 0.200 | -0.121 |
| 35 | Actions and interactions | showing, helping, giving, following, chasing, carrying, pointing_out, sending, waiting_for, throwing | -0.062 | 0.107 |
| 36 | Movement and direction | went, led, walked, snuck, leading, drove, rushed, traveled, came, scout | 0.111 | -0.034 |
| 37 | Education | lab, took_place, program, university, chemistry, science, psychology, history, episode, day_residue | 0.024 | 0.119 |
| 38 | Couple acts | met, meet, talked, talk, live, argue, spoke, stayed, lived, joined | 0.240 | 0.237 |
| 39 | Timing | wait, minutes, you, check, minute, leave, ticket, door, stay, an_hour | -0.048 | 0.096 |
| 40 | Housing locations | oak_valley, trailer, main_street, parked, east, home_avenue, driving, wilmerton, parking_lot, avenue | 0.105 | -0.218 |
| 41 | Questions and indefinite references | what, whatever, something_else, exactly_what, why, something, anything, whether, somebody, what_happened | -0.171 | 0.190 |
| 42 | Personal attributes and relationships | younger, blond, young, older, years_old, pregnant, boyfriend, naked, cousin, year_old | 0.012 | 0.014 |
| 43 | Negation and uncertainty | dont, didnt, not, never, doesnt, cant, did_not, nobody, nothing, never_did | -0.156 | 0.004 |
| 44 | State and condition | real_life, fact, terms, case, interest, waking_life, situation, manner, reality, life | 0.062 | 0.226 |

| 45 | Introductions | interviewing, introduced, introducing, introduces, sends, accompanied_by, welcomes, buys, lent, loaned | 0.233 | 0.054 |
|----|---------------|---------------------------------------------------------------------------------------------------------|-------|-------|
| 46 | Dream characters 2 | my_boyfriend, f, steve, mr, lou, judy, barbara, lee, ben_d, grandma_jane | -0.067 | 0.093 |
| 47 | Physical actions | noticed, saw, told, counted, picked, dropped, broke, indicated, picked_up, grabbed | -0.055 | 0.087 |
| 48 | Events and outings | coaches, auction, porn, pizzas, vendors, chili, functions, fragments, bands, world_war | 0.112 | -0.134 |
| 49 | Desire | makes, doesnt, does, loves, likes, doesnt_want, feels, doesnt_seem, demands, agrees | 0.077 | -0.029 |
| 50 | Dream characters and social acts | frank, marissa, zena, we, shake_hands, together, father_andrew, pedro, marissa_camden, chat | 0.412 | -0.059 |
| 51 | Daily events | visit, lunch, dinner, downstairs, shopping, upstairs, leave, home, supper, meet | -0.020 | 0.161 |
| 52 | Degree | very, how, too, so, fairly, extremely, quite, as, rather, enough | 0.058 | 0.148 |
| 53 | Qualities of journeys | long, fast, steep, far, slow, narrow, difficult, complicated, far_away, easy | -0.075 | -0.147 |
| 54 | Creative expression | play, dance, sing, ride, movie, playing, game, circle, stage, singing | 0.362 | -0.044 |
| 55 | Body postures | sitting, standing, seated, sat, bench, laying, lying, sat_down, couch, sit | 0.149 | -0.093 |
| 56 | Audiences | next, center, rest, audience, middle, front, group, side, now, front_row | 0.249 | 0.177 |
| 57 | Comfort | terribly, obvious, boring, efficient, painful, big_deal, fun, guilt, safer, strong | 0.068 | -0.234 |
| 58 | Dream recall | thought, think, guess, dreamed, dreamt_that, dreamt, thinking, week_ago, last_night, believe | -0.128 | 0.140 |
| 59 | Doubt | not, without, wonder_how, remain, wondering_if, nobody, otherwise, stronger_than, never, as | -0.012 | -0.076 |
| 60 | Ordering food and drink | buy, coffee, pay, eat, ice_cream, bought, drink, milk, ordered, give | 0.094 | 0.200 |
| 61 | Actions | jumped, pulled, ran, broke, flew, fell, took, got, stepped, dropped | -0.199 | -0.051 |
| 62 | Pronouns and relationships | me, us, him, each_other, directions, himself, them, his_wife, her, doctor_c | -0.046 | 0.301 |
| 63 | Household items | boxes, magazines, packages, items, plants, containers, hats, books, toys, coins | 0.185 | 0.013 |
| 64 | Grammatical structure | its, theres, looks, is, will_be, isnt, moves, gets, im, shes | -0.012 | -0.377 |
| 65 | Food containers | bottle, bag, tray, dish, milk, bowl, plastic, chocolate, meat, sheet | -0.034 | -0.063 |
| 66 | Clothes | clothes, shoes, pants, bed, knees, underwear, coat, bra, lap, shirt | -0.122 | -0.167 |
| 67 | Sensations | warm, weak, strong, tender, painful, gentle, powerful, strongly, intense, uncomfortable | -0.035 | -0.236 |
| 68 | Process | produces, versa, vague_ending, jean_k, effortlessly, crops, vice_versa, un, kay, freely | -0.024 | -0.280 |
| 69 | Proximity and position | or, leading, nearby, between, or_eight, perhaps, resembled, attached, divided_into, similar | -0.099 | -0.059 |
| 70 | Possession of relationships | my, my_fathers, my_mothers, franks, his, yearold, matthews, ss, daddys, jeremys | -0.111 | 0.069 |
| 71 | Bizarre dream experience | halloween, last_night, ______, celine_dion, wee_sing, crazy_dream, ________, bizarre_dream, dreamt_that, mini_disc | -0.055 | -0.034 |
| 72 | Colors | white, pink, blue, purple, brown, black, red, yellow, green, shiny | 0.128 | -0.321 |
| 73 | Initiating acts | ready, getting_ready, starting, going, back, decided, trying, while, supposed, tired | -0.151 | -0.010 |
| 74 | Work | test, word, report, information, list, book, job, program, email, project | 0.018 | 0.418 |
| 75 | Conflict | same_time, point, any_rate, moment, aimed, once, end, started_yelling, first, yelled | -0.299 | 0.465 |
| 76 | Personal attributes | bored, wealthy, thrilled, busy, horny, lazy, gay, poor, talented, successful | 0.134 | -0.173 |
| 77 | Time and movement | away, off, finally, until, after, away_from, eventually, immediately, before, over | -0.266 | 0.260 |
| 78 | Amounts of things | all_sorts, hundreds, all_kinds, lots, tons, whole_bunch, piles, stacks, full, different_types | 0.103 | -0.016 |
| 79 | Positive and silly experience | funny, stuff, weird, like, fun, stupid, crazy, interesting, silly, nice | 0.153 | -0.019 |
| 80 | Prepositions/adverbs of time | for, ago, after, already, since, later, during, before, worth, ordered | 0.122 | -0.056 |
| 81 | Outside movement | jump, fly, float, fall, crawl, slide, drop, dive, run, blow | -0.199 | -0.104 |
| 82 | Outdoor places (urban) | parking_lot, hallway, hall, street, woods, walking, mall, basement, streets, driveway | -0.041 | -0.106 |
| 83 | Social/professional interactions | question, if, answer, whom, answers, whether, who, questions, name, marry | 0.275 | 0.332 |
| 84 | Addresses | home_avenue, wilmerton, oak_valley, east, east_side, st_street, main_street, big_resort, nd_street, sliding | 0.045 | -0.281 |
| 85 | Intention | could, might, would, wouldnt, can, couldnt, should, wont, must, cant | 0.043 | -0.029 |
| 86 | Dream characters 3 | erin, williams, mack, charlene, johnson, donald, melissa, jeff, anne, kent | 0.068 | -0.093 |
| 87 | Dreaming | dream, dreamt_that, scene, morning, town, sudden, night, hall, street, school | -0.184 | 0.144 |
| 88 | Opening/closing | closed, locked, open, unlocked, ajar, opened, lock, sliding_glass, opening, unlock | -0.248 | -0.141 |
| 89 | Trying | trying, tried, try, kept_trying, managed, cant_seem, wanting, intending, able, being_able | -0.042 | 0.031 |
| 90 | Having to do something | id_better, id, im_gon, na, youll, theyd, youve, its_ok, weve_got, hed | -0.093 | -0.089 |
| 91 | Gateways | parking_lot, driveway, truck, garage, car, front_door, van, doorway, vehicle, yard | -0.108 | -0.006 |
| 92 | Quantity | lot, few, little_bit, bit, bigger, bunch_of, couple, little, more, shovel | -0.029 | -0.147 |
| 93 | Negative interactions with authority | phone, doctor, mrs, b, angrily, what_happened, god, crying, hello, question | -0.187 | 0.416 |
| 94 | Verbs | walk, step, go, crawl, sneak, peek, turn, dash, run, bend | -0.040 | -0.162 |
| 95 | Human interaction | sees, follows_me, watches, follows, decides, talks, likes, finds, says_hi, gets | 0.145 | -0.092 |
| 96 | Talking | say, do, saying, doing, say_oh, says, does, nod, wonder, happen | 0.225 | -0.139 |

| 97 | Travel problems | it, that, which, this, dangerous, pilot, plane, somehow, road, there | -0.123 | -0.075 |
|---|---|---|---|---|
| 98 | Transportation | train, road, boat, plane, traffic, highway, bus, freeway, land, motor | 0.091 | -0.094 |
| 99 | Military | they, tires, trucks, planes, guards, bees, birds, shells, prisoners, soldiers | 0.046 | -0.134 |
| 100 | Tools and weapons | hammer, rod, squirrel, rifle, bow, bat, shotgun, ball, chain, log | 0.061 | -0.084 |
| 101 | Water | sink, water, gas, plug, hose, slip, tub, urine, hole, faucet | -0.248 | -0.081 |
| 102 | School peers | may_be, is, from_brimson, has_been, are, meets, classmate_jerry, who_lives, lives, belongs | 0.183 | -0.084 |
| 103 | Seeming | seemed, appeared, didnt_seem, seems, doesnt_seem, expected, used, appears, pretended, seem | 0.058 | 0.017 |
| 104 | Movement with an other | follow, move, push, throw, catch, protect, escape, resist, pull, hide | 0.103 | 0.045 |
| 105 | Building features | ledge, roof, balcony, platform, staircase, wall, stairway, ramp, deck, slope | -0.050 | -0.241 |
| 106 | Readiness | away_from, ready, mad_at, away, unable, mad, safe, impatient, closer, tempted | -0.131 | 0.001 |
| 107 | City locations | city, area, country, small_town, building, mall, town, hotel, park, underground | 0.148 | -0.142 |
| 108 | Recollection | throughout, worst, actual, day_residue, during, events, whole, interpretation, recalled, exact | -0.009 | -0.016 |
| 109 | Groups | were, werent, various, both, different, arent, are, these, groups, mostly | 0.304 | 0.045 |
| 110 | Property descriptions | an_old, large, small, an_empty, nearby, victorian, huge, tiny, dining, small_square | 0.076 | -0.198 |
| 111 | Returning | back, returned, left, home, past, gone, return, locked, leave, front_door | -0.116 | 0.045 |
| 112 | Communicating 'nothing' | no, only, nothing, at_least, nobody, no_problem, dont_worry, im_sorry, sign, twice | 0.034 | -0.055 |
| 113 | Weather | bright, dark, clouds, sun, sky, light, thick, shiny, soft, beautiful | -0.119 | -0.304 |
| 114 | Outside features | be, had_been, whirlpool, vault, avoid_being, lion, sidewalks, creek, curb, platform | -0.174 | -0.036 |
| 115 | Extent and conjunctions | any, even, any_more, anymore, anything, though, much, particularly, but, too_much | -0.041 | 0.098 |
| 116 | Belief | wonder_why, agree, expect, believe, trust, feel_sad, respond, understand, recognize, intend | 0.074 | 0.134 |
| 117 | Movement into locations | out, through, into, inside, thru, outside, upstairs, downstairs, away, mail | -0.216 | 0.163 |
| 118 | Ordinal quantifiers | another, each, third, second, every, short, a, fourth, other, last | 0.045 | 0.129 |
| 119 | Quantities | three, several, groups, four, these, two, various, five, many, couple | 0.247 | 0.181 |
| 120 | Abrupt actions | walked_away, turned_around, woke_up, came, awoke, started, fell_asleep, went_downstairs, ran, got | -0.269 | 0.170 |
| 121 | Negative emotions | feel, feeling, fear, frustrated, intense, panic, felt, nervous, anxious, terribly | -0.327 | 0.060 |
| 122 | Take-out food | chips, soap, plastic, juice, strips, cardboard, dried, cans, bits, liquid | 0.077 | -0.200 |
| 123 | Object-oriented actions | fix, buy, pick_up, check, use, bring, write, find, send, remove | -0.017 | 0.128 |
| 124 | Expressions and | oh, mean, god, yes, hey, yourself, meaning, yours, yeah, better_than | 0.266 | -0.041 |
| | responses | | | |
| 125 | Body parts | arm, leg, grabbed, touched, shook, finger, held, hand, neck, hands | -0.112 | 0.011 |
| 126 | Man and weapons | he, himself, him, his, man, gun, bill, guy, knife, rifle | 0.097 | 0.216 |
| 127 | Dream characters 4 | whom, ben_d, carl, w, mary_kay, my_boyfriend, who, andy, ms, drama | -0.133 | 0.350 |
| 128 | Pronouns and contractions | hes, shes, theyre, im, youre, ive_been, its, weve_been, keeps, whos | -0.125 | -0.246 |
| 129 | Approaching | waved, sort, smiled, saw, approached, stood, met, reminds_me, watched, reminded_me | -0.069 | 0.029 |

**Table S2.** Coefficients and estimates from Partial Least Squares regression (PLS) and Linear Mixed-effect Models (LMM) for the effect of different network properties on dream valence and arousal. Results are shown for our selected model, with 130 topics. Significant p-values (<0.05) are highlighted in bold.

| | Valence | | | | Arousal | | | |
|---|---|---|---|---|---|---|---|---|
| Variable | PLS Coefficient | LMM estimate | PLS p-value | LMM p-value | PLS Coefficient | LMM estimate | PLS p-value | LMM p-value |
| Clustering | -0.0006 | -0.0109 | **0.031** | 0.085 | 0.0005 | -0.0043 | 0.190 | 0.659 |
| Cosine mod. | 0.0013 | 0.0172 | **<0.001** | **<0.001** | 0.0007 | 0.0253 | **0.022** | **<0.001** |
| Cosine path | 0.0005 | 0.0003 | **0.027** | 0.068 | -0.0002 | 0.0002 | 0.711 | 0.376 |
| Degree het. | -0.0018 | -0.0152 | **<0.001** | **<0.001** | 0.0022 | -0.0014 | **<0.001** | 0.758 |
| Efficiency | -0.0004 | -0.0099 | **0.020** | **0.032** | -0.0015 | -0.0162 | **<0.001** | **0.023** |
| Feedback loops | -0.0005 | -0.0189 | 0.068 | **0.004** | -0.0015 | -0.0256 | **0.007** | **0.012** |
| Gini | -0.0020 | -0.0345 | **<0.001** | **<0.001** | 0.0020 | -0.0140 | **<0.001** | 0.150 |
| Mean coherence | 0.0054 | 0.0486 | **<0.001** | **<0.001** | -0.0327 | -0.5768 | **<0.001** | **<0.001** |
| Modularity | 0.0012 | 0.0164 | **<0.001** | **0.001** | 0.0002 | 0.0170 | 0.711 | **0.023** |
| Nodes | -0.0011 | -0.0001 | **<0.001** | 0.061 | 0.0035 | 0.0000 | **<0.001** | 0.927 |
| Std. Coherence | 0.0069 | 0.1982 | **<0.001** | **<0.001** | -0.0258 | -0.7265 | **<0.001** | **<0.001** |
| Topic path | 0.0005 | 0.0003 | **0.027** | 0.069 | -0.0001 | 0.0002 | 0.711 | 0.362 |
| Transitivity | -0.0008 | -0.0174 | **0.006** | **<0.001** | -0.0010 | -0.0307 | **0.029** | **0.002** |

**Table 1.** Definitions and interpretations of network properties analyzed in dream topic networks. Network illustrations indicate each property in a low and high state. Nodes (circles) represent topics and edges (arrows) the transitions between successive topics. Longer edges indicate that connected topics are more semantically dissimilar (further apart in our embedding), whilst shorter edges indicate that connected topics are more semantically similar (closer together in our embedding). Yellow networks in the high column indicate a property predicted to be positively associated with valence whilst yellow networks in the low column indicate a property predicted to be negatively associated with valence.

| Property | Definition | Interpretation | Low | High |
|---|---|---|---|---|
| Mean coherence | The average semantic similarity of connected topics | Increases with narrative disorder | 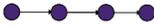 | 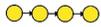 |
| Standard deviation of coherence | The variation in the semantic similarity of connected topics | Increases with transitions between distinct domains in semantic space | 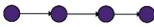 | 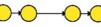 |
| Modularity | The extent to which a network is structured into distinct communities, with edge weights given by the frequency of topical transitions | Increases with narrative structure | 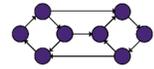 | 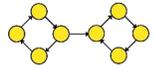 |
| Cosine modularity | The extent to which a network is structured into distinct communities, with edge weights given by the cosine similatly between connecting topics | Increases with narrative structure | 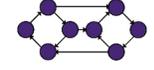 | 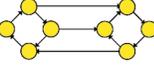 |
| Topic path | The minimum number of steps between the first (S) and last (E) topic in a network | Increases with narrative linearity | 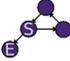 | 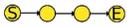 |
| Cosine path | The shortest inverse-cosine weighted path between the first (S) and last (E) topic in a network | Increases with narrative linearity | 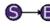 | 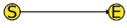 |
| Efficiency | The reciprocal of the average-shortest path length between nodes in a network | Decreases with narrative linearity | 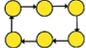 | 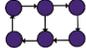 |
| Feedback-loops | The proportion of connected nodes with bidirectional edges, amongst those that are connected | Increases with rumination | 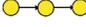 | 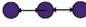 |
| Clustering | The tendency of neighboring nodes to form fully connected sets | Increases with rumination | 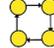 | 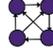 |
| Transitivity | The proportion of triangles (loops between three nodes), amongst those that are possible | Increases with rumination | 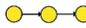 | 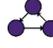 |
| Degree heterogeneity | The coefficient of variation in the degree of nodes in a network | Increases with narrative disorder, topic dominance | 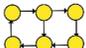 | 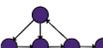 |
| Gini coefficient | The inequality in the degree distribution in a network | Increases with topic dominance | 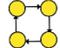 | 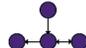 |
| Nodes | The overall number of unique topics in a network | Increases with narrative size | 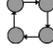 | 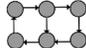 |